\documentclass[12pt,letterpaper]{article}
\usepackage{amsfonts}
\usepackage{amsmath}
\usepackage{amsthm}
\usepackage{amssymb}
\usepackage{geometry}
\usepackage{xcolor}
\usepackage{graphicx}
\usepackage[onehalfspacing]{setspace}
\newtheorem{assumption}{Assumption}
\newtheorem{theorem}{Theorem}

\newtheorem{proposition}{Proposition}
\newtheorem{remark}{Remark}
\usepackage[authoryear,round,longnamesfirst]{natbib}

\begin{document}
\onehalfspacing
\title{Non-Robustness of the Cluster-Robust Inference: \\ with a Proposal of a New Robust Method\thanks{This paper has been merged into arXiv:2308.10138v6 (``Genuinely Robust Inference for Clustered Data'') as of January 27, 2025. Please refer to the merged paper. First arXiv date: October 31, 2022. We benefited from comments by and discussions with A. Colin Cameron, Bruce Hansen, Seojeong (Jay) Lee, Ulrich M\text{\"u}ller, and Max Tabord-Meehan. We would like to thank Georgina Cisneros, Pratap Khattri, Kalika Likhi, Perry Sharaf, Hong Xu and Siyuan Xu for excellent research assistance. All remaining errors are ours.}}
\author{Yuya Sasaki\thanks{
Associate professor of economics, Vanderbilt University. Email: \texttt{yuya.sasaki@vanderbilt.edu}} 
\ and 
Yulong Wang\thanks{Assistant professor of economics, Syracuse University. Email: \texttt{ywang402@syr.edu}}}
\date{}
\maketitle

\begin{abstract}
\setlength{\baselineskip}{6mm}
The conventional cluster-robust (CR) standard errors may not be robust.
They are vulnerable to data that contain a small number of large clusters.
When a researcher uses the 51 states in the U.S. as clusters, the largest cluster (California) consists of about 10\% of the total sample.
Such a case in fact violates the assumptions under which the widely used CR methods are guaranteed to work.
We formally show that the conventional CR methods fail if the distribution of cluster sizes follows a power law with exponent less than two.
Besides the example of 51 state clusters, some examples are drawn from a list of recent original research articles published in a top journal.
In light of these negative results about the existing CR methods, we propose a weighted CR (WCR) method as a simple fix.
Simulation studies support our arguments that the WCR method is robust while the conventional CR methods are not.

{\small { \ \ \newline
}
}
\end{abstract}

\setlength{\baselineskip}{7.45mm}
\newpage
%%%%%%%%%%%%%%%%%%%%%%%%%%%%
\section{Introduction}\label{sec:introduction}
%%%%%%%%%%%%%%%%%%%%%%%%%%%%

Cluster-robust (CR) standard errors account for within-cluster correlations.
Such correlations often arise by construction within an industry \citep{He1998} or within a state \citep{BeDuMu2004}, to list a couple of the earliest examples.
Today, even if a model may not induce cluster dependence by construction, applying CR methods by observable group identifiers is quite common in practice.

While CR methods have been used for decades, ``it is only recently that theoretical foundations for the use of these methods in many empirically relevant situations have been developed'' \citep{MaNiWe2022}.
The initial theory fixes the (maximum) cluster size $N_g$ to a small number and assumes some large number $G$ of clusters \citep[][]{Wh84,LiZe86,Ar87}.
More recent theory \citep[][]{DjMaNi19,HaLe19,Ha22} allows for large $N_g$, but still imposes the restriction $\sup_g N^2_g / N \rightarrow 0$ of vanishing maximum cluster size relative to the whole sample size $N = \sum_{g=1}^G N_g$ as $G \rightarrow \infty$, for establishing the convergence rate and asymptotic distribution of the estimator. 

These conventional assumptions may not be always satisfied.
They do not hold for certain important situations, such as the case in which a researcher uses the 51 states in the U.S. as clusters.
As documented later in this article, the distribution of such cluster sizes $N_g$ appears to follow the power law.
Consequently, $\sup_g N^2_g / N \rightarrow 0$ will not hold.
To fix ideas, think of $\sup_g N_g^2 / N \gg \sup_g N_g / N \approx 0.1 \gg 0$ for the state of California as an outlier in terms of the cluster size in typical survey data. 

We show that the assumption of bounded second moments for the CR score fails if the distribution of $N_g$ follows the power law with the tail exponent less than two.
For the example of using the 51 states as clusters, unfortunately, we cannot rule out the possibility of the tail exponent being less than two.
Furthermore, we show that the conventional assumption $\sup_g N_g^2/N \rightarrow 0$ fails too.
Therefore, we cannot recommend using the conventional CR methods for such data.
We highlight several examples from a list of original research articles recently published in a top journal.

%In light of these negative findings concerning the conventional CR methods, we propose a simple fix.
We propose a simple fix in light of these adverse findings concerning the conventional CR methods.
Specifically, we propose a weighted CR (WCR) method to suppress heavy tails in the distributions of cluster sums of scores.
We provide supporting theories to guarantee that this proposed method works.
Simulation studies demonstrate that our proposed WCR method achieves significantly more accurate coverage than the conventional CR methods, thus supporting our arguments that the WCR method is robust while the conventional CR methods are not.

%%%%%%%%%%%%%%%%%%%%%%%%%%%%
\section{Relation to the Literature}\label{sec:literature}
%%%%%%%%%%%%%%%%%%%%%%%%%%%%

The initial theory \citep*[][]{Wh84,LiZe86,Ar87} for CR methods assumes small cluster sizes $N_g$ as $G \rightarrow \infty$.
It is implemented by the `\texttt{cluster()}' option or the `\texttt{vce(cluster)}' option by Stata, and is used by almost all, if not all, empirical papers that report CR standard errors.
It has been known that a large cluster size $N_g$ in data could lead to a large CR standard error \citep*[e.g.,][p. 324]{CaMi15}.

More recently, \citet*{Ha07}, \citet*{IbMu10,IbMu16}, \citet*{BeCoHa11}, and \citet*{CaSaSh21} propose alternative theory to accommodate large $N_g$ while $G$ is fixed.
Their frameworks differ from ours in a few aspects.
First, they exploit within-cluster central limit theorem (CLT) which is suitable for panel data, while we focus on the cross-sectional settings which accommodate arbitrary within-cluster dependence as opposed to weak dependence.
Second, they consider sequences of increasing cluster sizes $N_g$ \textit{for all} $g$ which is suitable for long panels, while we treat $N_g$ as random variables drawn from a widely supported distribution to accommodate both small clusters (like the state of Wyoming) and large clusters (like the state of California) in cross-sectional settings.
Third, they consider small $G$ while we consider large $G$.

As such, more closely related to this article are the recent developments by \citet*[][]{DjMaNi19}, \citet*{HaLe19} and \citet*{Ha22}.
Both their frameworks and our framework consider large $G$ with arbitrary within-cluster dependence as opposed to weak dependence.
The asymptotic theories in these papers essentially require $\sup_g N^2_g/N \rightarrow 0$.
As mentioned in Section \ref{sec:introduction}, however, we have $\sup_g N^2_g / N \gg \sup_g N_g / N \approx 0.1 \gg 0$ for the state clusters in typical survey data in the U.S.
More generally, we find that the assumption of $\sup_g N^2_g/N \rightarrow 0$ fails if the distribution of $N_g$ follows the power law, as is the case with the state clusters among others.
Consequently, we propose the alternative WCR method, accommodating cases with $\sup_g N^2_g/N \not\rightarrow 0$. 
With this said, we want to stress that our framework with the power law does not necessarily nest those of these preceding papers.

%Our proposal is to weight clusters by $1/N_g$.
%%In contrast, \citet[][p. 137]{Wo03} proposes to weight clusters by $N_g/\hat\sigma_g$ where $\hat\sigma_g$ denotes the cluster sample variance.
%%The underlying settings and motivations are different between these contrasting proposals of the weights.
%Under substantially different settings, \citet[][p. 137]{Wo03} proposes to weight clusters by $N_g/\hat\sigma^2_g$ where $\hat\sigma^2_g$ denotes the cluster sample variance. 
%The motivation for such weight is to improve efficiency when $N_g$ is large and $\hat\sigma^2_g$ consistently estimates the true variance. 
%In constrast, our framework allows $N_g$ to be random and hence very small in some groups. 
%Consistent estimation of the variance in each cluster is out of the question. 
%Moreoever, \citet{Wo03} suggests his weight for efficiency under small $G$, while we suggest our weight to satisfy the bounded second moment condition for scores for valid CR inference under large $G$.

Furthermore, it has been known that the conventional CR standard errors are biased downward, and the jackknife estimator exhibits a better performance \citep[e.g.,][]{Ha2022,MaNiWe2022fast,MaNiWe2022leverage}.
When the cluster size $N_g$ follows a power law, however, the self-normalized means may converge to a non-Gaussian limit in general, and hence even the jackknife is not guaranteed to work in such cases.
That being said, we suggest to combine our proposed WCR method with the jackknife variance estimation.
Specifically, we suggest the HC3 estimator \citep{MaWh85} with the \textit{weighted} cluster sum as an effective unit of observation.

Finally, this paper is also closely related to the recent literature on cluster randomized experiments \citep*[e.g.][]{BaLiShTa2022,BuCaShTa2022,CrToVa2022} in a couple of ways, even though the main objectives in this literature are different from ours.
First, while the existing literature discussed thus far treats $N_g$ as deterministic sizes, this new literature treats $N_g$ as stochastic sizes similarly to our paper.
While this recent literature assumes $\sup_g N_g^2/N \stackrel{p}{\rightarrow} 0$ (and equivalently $\mathbb{E}[N_g^2]<\infty$), however, we allow $N_g$ to be drawn from heavy-tailed distributions so $\sup_g N_g^2/N \stackrel{p}{\rightarrow} 0$ need not hold.
Second, our proposed WCR estimator encompasses one of the estimators proposed by \citet*{BuCaShTa2022}.
This implies that their estimator in fact works under our assumption as well, and consequently, their approach does not require $\sup_g N_g^2/N \stackrel{p}{\rightarrow} 0$ under the set of our assumptions.

Besides those discussed above, there are many important papers in the extensive literature of CR methods.
We refer readers to the comprehensive surveys by \citet*{CaMi15,CaMi22} and \citet*{MaNiWe2022}.

\section{Review of the Cluster-Robust Inference}
%%%%%%%%%%%%%%%%%%%%%%%%%%%%
Often available for empirical research is a sample of $G$ clusters, where the $g$-th cluster consists of $N_g \in \mathbb{N}$ observations for each $g \in \{1,\cdots,G\}$.
A common assumption in this setting is that observations within a cluster may be arbitrarily correlated, but they are assumed to be independent across clusters.
For instance, the fifty-one states are often treated as $G=51$ clusters based on a random sample drawn from the population in the U.S., allowing state-specific factors to induce dependence among observations within each state possibly.  

To fix ideas, suppose that a researcher uses a clustered sample $\{\{(Y_{gi},X_{gi}')'\}_{i=1}^{N_g}\}_{g=1}^G$ to study the linear model
\begin{align*}
Y_{gi} = X_{gi}'\theta + U_{gi},
\qquad
\mathbb{E}[U_{g}|X_{g}]=0,
\end{align*}
where $U_g = (U_{g1},\cdots,U_{gN_g})'$ and $X_g = (X_{g1},\cdots,X_{gN_g})'$ for each $g \in \{1,\cdots,G\}$.
Also write $Y_g = (Y_{g1},\cdots,Y_{gN_g})'$ for each $g \in \{1,\cdots,G\}$.
Then, the ordinary least squares (OLS) estimator for the parameter vector $\theta$ under the cluster sampling takes the form of
\begin{align}\label{eq:ols}
\widehat\theta 
= 
\left(\sum_{g=1}^G \sum_{i=1}^{N_g} X_{gi} X_{gi}'\right)^{-1} \left(\sum_{g=1}^G \sum_{i=1}^{N_g} X_{gi} Y_{gi}\right)
= 
\left(\sum_{g=1}^G X_g'X_g\right)^{-1} \left(\sum_{g=1}^G X_g'Y_g\right).
\end{align}

Furthermore, commonly employed cluster-robust (CR) variance estimators take the form of
\begin{align}\label{eq:cr}
\widehat V_{\hat\theta}^{\text{CR}} = a_n \left(\sum_{g=1}^G X_g'X_g\right)^{-1} \left(\sum_{g=1}^G \widehat S_g \widehat S_g' \right) \left(\sum_{g=1}^G X_g'X_g\right)^{-1},
\end{align}
where $a_n \rightarrow 1$ is a suitable finite-sample adjustment and
$
\widehat S_g = \sum_{i = 1}^{N_g} X_{gi} \widehat U_{gi}
$
with
$
\widehat U_{gi} = Y_{gi} - X_{gi}'\widehat\theta.
$
In particular, a majority of empirical economics papers is based on the `\texttt{cluster()}' option or the `\texttt{vce(cluster)}' option in Stata, which uses \eqref{eq:ols} and \eqref{eq:cr} with the finite-sample adjustment
\begin{align*}
a_n = \left(\frac{\sum_{g=1}^G N_g-1}{\sum_{g=1}^G N_g-\dim\{X_{gi}\}}\right) \left(\frac{G}{G-1}\right).
\end{align*}
% where $p$ denotes the dimension of $X_{gi}$.
Note that this adjustment factor satisfies $a_n \rightarrow 1$ as $G \rightarrow \infty$.

Also used is the jackknife variance estimator defined by
\begin{align}\label{eq:crjack}
\widehat V_{\hat\theta}^{\text{CR,JACK}} = \sum_{g=1}^G \left(\widehat\theta_{-g}-\widehat\theta\right) \left(\widehat\theta_{-g}-\widehat\theta\right)',
\end{align}
where $\widehat\theta_{-g}$ denotes the leave-one-cluster-out estimator defined by
$$
\widehat\theta_{-g} = \left(\sum_{h \neq g} X_h'X_h\right)^{-1} \left(\sum_{h \neq g} X_h'Y_h\right).
$$

%%%%%%%%%%%%%%%%%%%%%%%%%%%%
\section{A Pitfall in the Cluster-Robust Inference}\label{sec:pitfall}
%%%%%%%%%%%%%%%%%%%%%%%%%%%%

While they provide robustness to cluster dependence, the CR methods are vulnerable to common cross-sectional situations in which there are a small number of large clusters.

The standard econometric theory to guarantee that the OLS estimator \eqref{eq:ols} and the CR variance estimator \eqref{eq:cr} behave well under the cluster sampling are based on the asymptotic property
\begin{align}\label{eq:asymptotic}
\sqrt{G} \left( \widehat\theta - \theta \right) 
= 
\Big( \underbrace{ \frac{1}{G} \sum_{g=1}^G \Xi_g}_{\stackrel{p}{\longrightarrow} \ Q} \Big)^{-1}
\Big( \underbrace{ \frac{1}{\sqrt{G}} \sum_{g=1}^G S_g }_{\stackrel{d}{\longrightarrow} \mathcal{N}(0,V)} \Big)
\stackrel{d}{\longrightarrow}
\mathcal{N}(0,Q^{-1}VQ^{-1})
\end{align}
as $G \rightarrow \infty$,
where $\stackrel{p}{\rightarrow}$ stands for convergence in probability, $\stackrel{d}{\rightarrow}$ stands for convergence in distribution, and
\begin{align*}
&Q = \mathbb{E}[\Xi_g],
&&\Xi_g = \sum_{i=1}^{N_g} X_{gi}X_{gi}',
\\
&V = \text{Var}[S_g],
&&S_g = \sum_{i = 1}^{N_g} X_{gi} U_{gi}.
\end{align*}

Indeed, the cluster-robust approach allows for robustness against within-cluster dependence.
However, this robustness is not cost-free.
Namely, suitable moments of $\Xi_g$ and $S_g$ need to be finite for the weak law of large numbers and the central limit theorem to be invoked in \eqref{eq:asymptotic}.
Specifically, the asymptotic normality \eqref{eq:asymptotic} and the validity of the CR variance estimator \eqref{eq:cr} under the cluster sampling require $S_g$ to have finite second moments.
If the second moment of $S_g$ does not exist, then $\sqrt{G}\left(\widehat\theta-\theta\right)$ diverges and hence would not converge in distribution as in \eqref{eq:asymptotic}.
However, we are going to argue that this bounded second moment condition for the cluster score $S_g$ can fail under the cluster sampling \textit{even if} the second moment of the individual's score $X_{gi}U_{gi}$ were finite.

For ease of illustration, we first introduce a few definitions and notations.
For a given $j \in \{1,\cdots,\dim\{X_{gi}\}\}$, let $\Sigma_g$ and $Z_{gi}$ be short-hand notations for the $j$-th coordinate of $S_g$ and the $j$-th coordinate of $X_{gi}U_{gi}$, respectively -- we omit the dependence on $j$ in these notations for brevity.
For any distribution function $F$, we say that $F$ is regularly varying (RV) at infinity if it satisfies the following property that
\begin{equation*}
\frac{1-F(xt)}{1-F(t)}\rightarrow x^{-\alpha} \text{ as } t \rightarrow \infty,
\end{equation*}
for any $x>0$ and some constant $\alpha>0$. 
The constant $\alpha$ is referred to as the tail exponent, which measures the tail heaviness of $F$. 
Let $F$ denote the marginal distribution of $Z_{gi}$.
For each $n \in \{2,3,\cdots\}$, let $C^n$ denote the copula that characterizes the joint distribution of $(Z_{g1},\dots,Z_{gn})$ such that 
$$
\mathbb{P}\left( Z_{g1}\leq z_1,\dots,Z_{gn} \leq z_n \right) =
C^n(F(z_1),\dots,F(z_n)).
$$
With these definitions and notations, we make the following set of assumptions.

\begin{assumption}\label{a:cluster}
The following conditions are satisfied.
\begin{enumerate}
\item\label{a:cluster:rv} 
$\{Z_{gi}\}_{gi}$ is identically distributed with CDF $F$, which is RV at infinity with coefficient $\alpha > 1$.

\item\label{a:cluster:copula}
The copula density  $c(u_{1},\dots ,u_{n})={\partial ^{n}C(u_{1},\dots ,u_{n})}/{\partial u_{1}\cdots \partial u_{n}}$ exists and is uniformly bounded.

\item\label{a:cluster:indep}
$N_g$ is independent of $(Z_{g1},Z_{g2},....)$ and its distribution $H$ is RV at infinity with coefficient $\beta > 1$.
\end{enumerate}
\end{assumption}

We provide some discussions about Assumption \ref{a:cluster}. 
Assumption \ref{a:cluster}.\ref{a:cluster:rv} requires that the CDF of $Z_{gi}$ implies a finite mean and its tail is regularly varying. 
The regularly varying tail is mild and satisfied by many commonly used heavy-tailed distributions such as Pareto, Student-t, Cauchy, F distributions. 
Without loss of generality, we focus on the right tail and consider that $Z_{gi}$ is non-negative. 
Otherwise, our conclusion still holds provided \citep[cf.][p. 659]{ZhShWe09}
\begin{equation*}
\lim_{z\rightarrow\infty}\frac{\mathbb{P}(Z_{gi}<-z)}{\mathbb{P}(Z_{gi}>z)}=0.
\end{equation*} 
%{\color{red}{[What do you mean by the denominator, $Z_{gi}>z$? It is an inequality, not a number.]}}
Given the regularly varying tail, we can elegantly characterize the moment conditions \citep[e.g.,][Proposition 1.3.2]{mikosch1999regular} as
\begin{align*}
\mathbb{E}[|Z_{gi}|^r]&=\infty \text{ for any } r<\alpha \\
\mathbb{E}[|Z_{gi}|^r]&<\infty \text{ for any } r>\alpha.
\end{align*}
When $Z_{gi}$ has bounded infinite-order moments (such as Gaussian distribution) or even has a compact support, the corresponding tail exponent can be considered to be arbitrarily large. 
See, for example, \citet[][Chpater 1]{de07} for a more precise statement in terms of $1/\alpha$. 
These distributions are commonly referred to as thin-tailed distributions, which are still covered by our conclusion. 
For conciseness, we suppress them in this assumption but still implement them in the simulation studies in Section \ref{sec:simulation}. 
Assumption \ref{a:cluster}.\ref{a:cluster:copula} allows for dependence among $Z_{g1},\dots,Z_{gN_g}$ within each cluster $g$. 
This condition allows for very general formats of correlations within each cluster. 

Assumption \ref{a:cluster}.\ref{a:cluster:indep} requires that the distribution of the cluster size $N_g$ is also regularly varying. 
This condition is key to our results and is plausibly satisfied as shown in the following section.
Since $N_g$ takes integer values, we may consider it as the integer part of some continuous random variable, say $N^*_g$ whose tail is regularly varying. 
Given that $N^*_g-1 \leq N_g\leq N^*_g$, we have that $\mathbb{P}(N_g>y)\sim \mathbb{P}(N^*_g>y)$ as $y\rightarrow\infty$ and hence continue with $N_g$ for conciseness. 
In some applications, the requirement of independent cluster sizes in Assumption \ref{a:cluster}.\ref{a:cluster:indep} may be too strong. 
\citet*{BuCaShTa2022} for instance consider non-independent cluster sizes for cluster randomized experiments.
For later results, we can relax this requirement at the cost of strengthening Assumption \ref{a:cluster}.\ref{a:cluster:copula} -- see Remarks \ref{remark:relaxing} and \ref{remark:relaxing2} ahead for details.

Suppose that $\beta < \alpha$ under Assumption \ref{a:cluster}.\ref{a:cluster:rv} and \ref{a:cluster}.\ref{a:cluster:indep}.
In other words, the tail of the distribution of $N_g$ dominates that of $Z_{gi}$.
In this case, the tail heaviness of the summation $\Sigma_g=\sum_{i=1}^{N_g}Z_{gi}$ is dictated by that of $N_g$. 
Therefore, \textit{even if} $Z_{gi}$ has a finite $r$-th moment for $r<\alpha $, the $r$-th moment of $\Sigma_g$ might still be infinite if $r>\beta $. 
The following theorem formalizes this argument.

\begin{theorem}
\label{prop:tail} 
Suppose Assumption \ref{a:cluster} holds with $\beta<2$. Then,
\[
\frac{\sup_g N_g^2}{N}
\rightarrow \infty \text{ with probability approaching 1 as } G\rightarrow \infty. 
\]
Moreover, if $\beta < \alpha$, then for any $z>0$, 
\[
\frac{\mathbb{P}(\Sigma_g > zt)}{\mathbb{P}(\Sigma_g > t)} = z^{-\beta} \text{ as } t\rightarrow \infty.
\]
\end{theorem}

\noindent
We provide a proof in Appendix \ref{sec:prop:tail}.

Under i.i.d. sampling, the `robust' standard errors (e.g., those based on Eicker-Huber-White variance estimator or the \texttt{robust} option in Stata) exist and behave well if $Z_{gi}$ has a bounded second moment.
Under cluster sampling, however, the cluster-`robust' standard errors (i.e., those based on \eqref{eq:cr} or the \texttt{cluster()}/\texttt{vce(cluster)} options in Stata) may fail to exist even if $Z_{gi}$ has a bounded second moment.
The first part of Theorem \ref{prop:tail} shows that the assumption of $\sup_g N_g^2/N \rightarrow 0$ which has been imposed by even the recent literature (cf. Section \ref{sec:literature}) on cluster-robust inference is also implausible under the power law.\footnote{\citet{DjMaNi19} discuss the scenario in which $\sup_g N_g^2/N \rightarrow 0$ is relaxed to $\sup_g N_g/N \rightarrow 0$ under some additional assumptions. We find that such relaxation could be violated under our framework with a random $N_g$. See Appendix \ref{sec:appendix:self_normalized} for details. }
The second part of Theorem \ref{prop:tail} further implies that $\Sigma_g$ may have an infinite second moment \textit{even if} $Z_{gi}$ has a bounded second moment.
This occurs when $\beta < 2 < \alpha$.

%%%%%%%%%%%%%%%%%%%%%%%%%%%%
\section{Example: Problem with the 51 States as Custers}\label{sec:states}
%%%%%%%%%%%%%%%%%%%%%%%%%%%%

A natural question now is whether the distribution of cluster sizes $N_g$ has a heavy tail with tail index $\beta < 2$.
An answer will of course depend on data and the context of empirical research.
Let us focus on one of the most common situations of empirical economic analyses with clustering.
Specifically, suppose that a researcher clusters a sample of individuals in the U.S. Panel Study of Income Dynamics (PSID) by the 51 states.
We extract the sample of all the male individuals from the most recent wave of 2019.
This sample consists of $4808$ observations across $G=51$ states.

\begin{figure}[h]
\centering
\includegraphics[width=10.5cm]{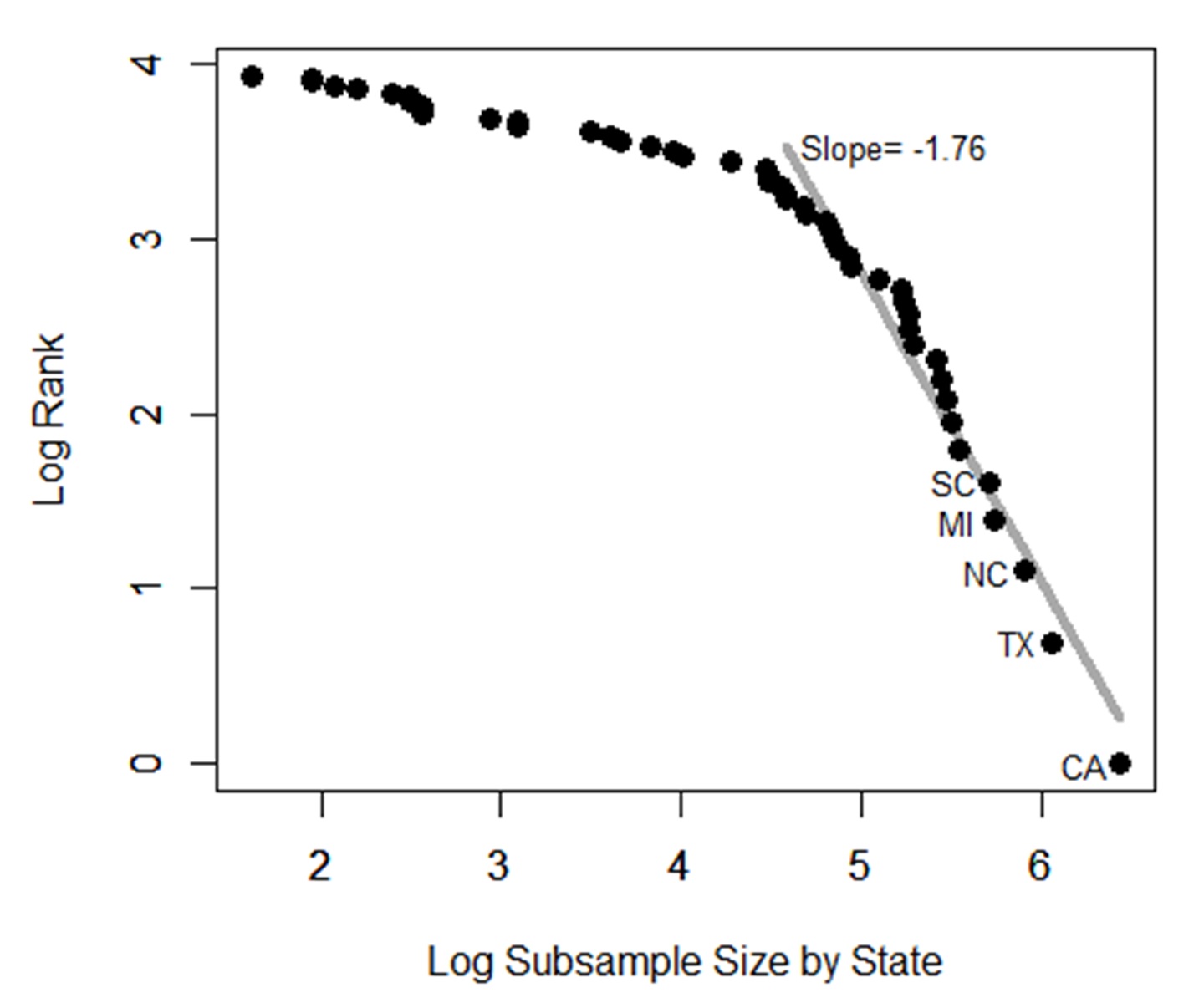}
\caption{The log-log plot for the subsamples of male individuals by state clusters in the U.S. PSID wave of 2019.}${}$
\label{fig:psid_loglog}
\end{figure}

In this data set, the five largest clusters have their sizes of $N_{4} = 629$ (California), $N_{42} = 430$ (Texas), $N_{32} = 366$ (North Carolina), $N_{21} = 310$ (Michigan), and $N_{39} = 304$ (South Carolina).
Figure \ref{fig:psid_loglog} plots the logarithm of the rank of $N_g$ (in descending order) against the logarithm of $N_g$ for all the 51 clusters.
The gray line indicates the fitted line using the largest 25 observations, i.e., 50\% of the 51 states.
Note that these largest 25 states follow the line of slope $-1.76$, while the remaining 26 states follow another line of a less steep slope.
This shape of the log-log plot implies that the distribution of the cluster sizes $N_g$ is approximately asymmetric double Pareto.
The slope of $-1.76$ can be interpreted as the negative value of the Pareto exponent $\beta$ in the right tail of the distribution $H$ of $N_g$.
Hence, the first part of Theorem \ref{prop:tail} implies that the conventional assumption $\sup_g N_g^2/N \rightarrow 0$ is likely to fail.

\begin{figure}[h]
\centering
\includegraphics[width=11cm]{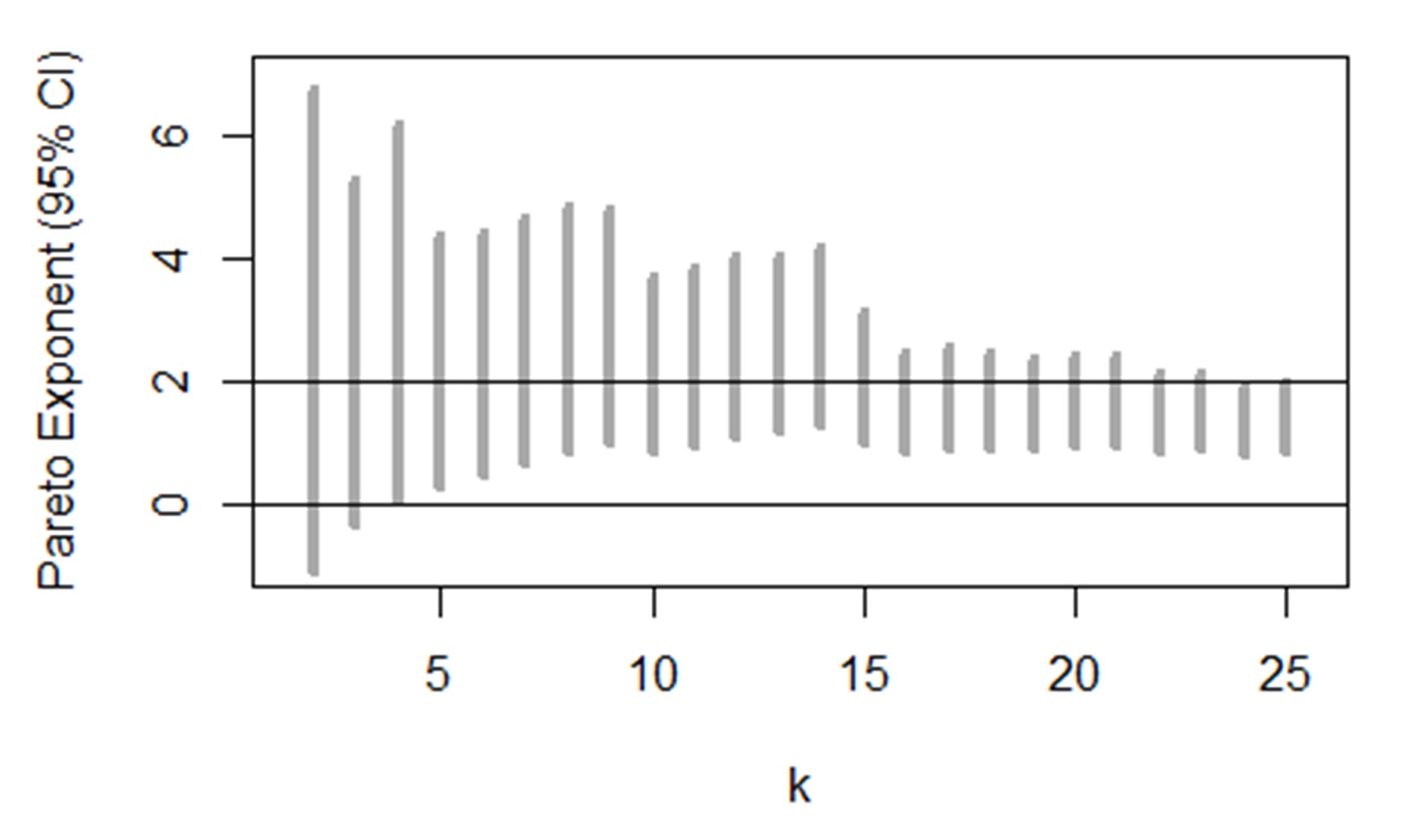}
\caption{The Hill plot for the largest 25 (50\%) of the 51 state clusters in the U.S. PSID wave of 2019: men.}${}$
\label{fig:psid_hill}
\end{figure}

From this graphical analysis, it seems quite possible that $\beta < 2$.
We now resort to a more formal and widely practiced test to reinforce this heuristic observation. 
Figure \ref{fig:psid_hill} displays the so-called Hill plot \citep[e.g.,][]{DrReDe2000}, consisting of 95\% confidence intervals for the Parteto exponent $\beta$ of the distribution of $N_g$ using the largest $k$ order statistics for $k \in \{2,\cdots,25\}$.
Observe that we cannot rule out the possibility of $\beta < 2$ for any $k$.

Recall from Theorem \ref{prop:tail} that the Pareto exponent $\beta$ of the distribution $H$ of $N_g$ dictates the heaviness of the distribution of the coordinates $\Sigma_g$ of the score $S_g$.
If $\beta < 2$, then the asymptotic convergence \eqref{eq:asymptotic} fails \textit{even if} $\alpha > 2$ holds for each coordinate $Z_{gi}$ of the score $X_{gi}U_{gi}$.
In light of the conclusion from the previous two paragraphs, we cannot rule out $\beta < 2$.
Therefore, it follows that the widely employed CR variance estimator \eqref{eq:asymptotic} may lead to misleading inference.
In summary, we do not recommend the common practice of using the CR standard error with the 51 states as clusters for this data set from the U.S. PSID.

%%%%%%%%%%%%%%%%%%%%%%%%%%%%
\section{Examples from \textit{Econometrica}, 2020}
%%%%%%%%%%%%%%%%%%%%%%%%%%%%

We highlight several examples based on recent original research articles published in 2020 in \textit{Econometrica} that use cluster-robust methods of inference for which we cannot rule out the possibility of $\beta < 2$.
Namely we showcase in point with
state clusters in the U.S. \citep{BuHaTiVo20},
region clusters in Russia \citep{EnMaPe20}, and
NGO branch clusters \citep{AlBaBaBuRaSuVi20}.

We would like to emphasize that the main point of this section is not to criticize a specific list of papers -- the authors of these papers are in fact great for allowing us to replicate their analyses more easily than many others.
Our selection from one journal from only one recent year reflects only the tip of the iceberg -- furthermore, there are even more articles published in \textit{Econometrica} and other journals for which data are unavailable to us or replication was difficult for us.
We suspect that an enormous number of empirical research papers from a wide variety of journals and from a long history of the literature are in fact subject to this problem of non-robustness.

%%%%%%%%%%%%%%%%%%%%%%%%%%%%
\subsection{State Clusters in the U.S. \citep{BuHaTiVo20}}
%%%%%%%%%%%%%%%%%%%%%%%%%%%%

\begin{figure}[h]
\vspace{1cm}\centering
\includegraphics[width=11cm]{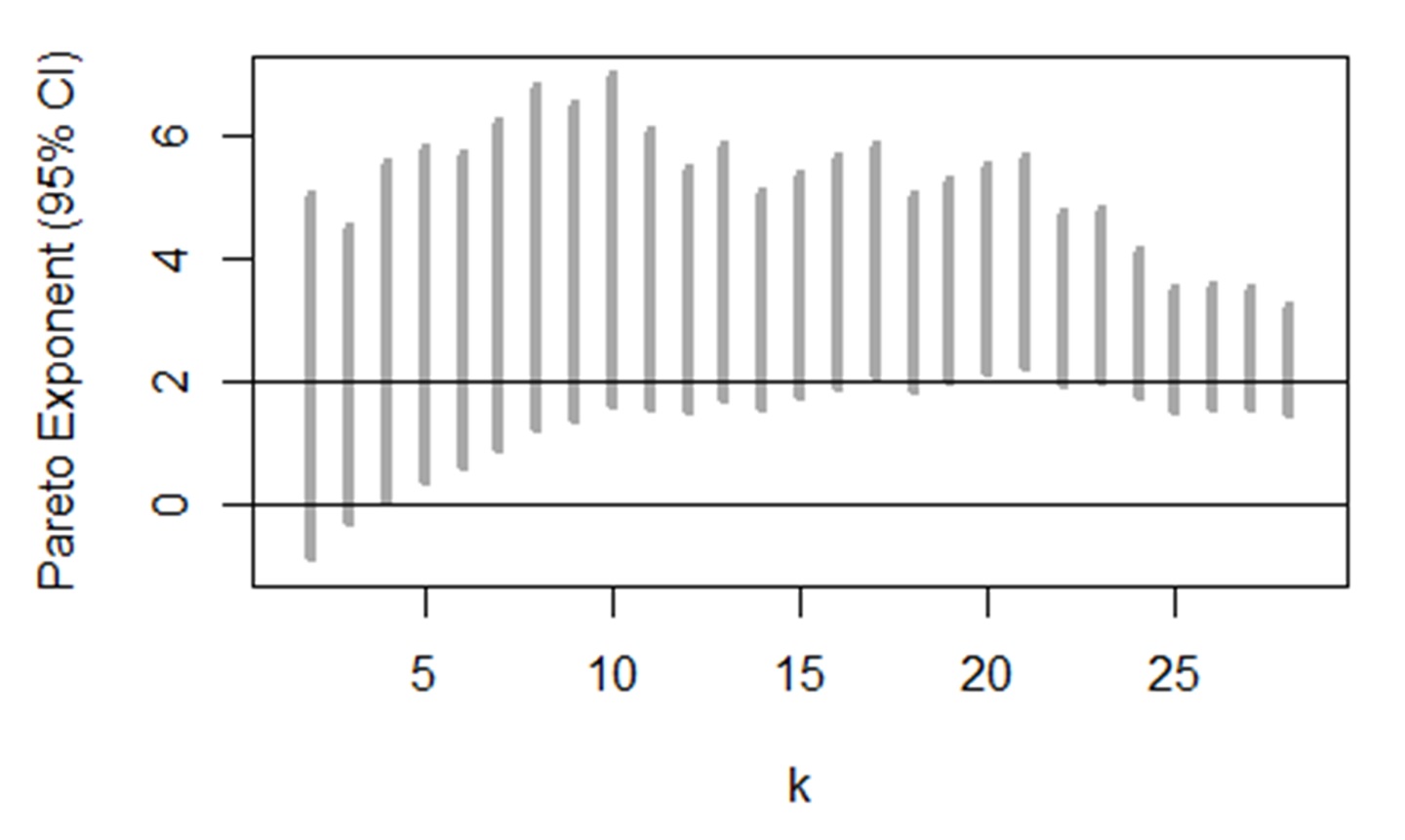}
\caption{The Hill plot for the largest 28 (50\%) of the 56 state clusters for \citep{BuHaTiVo20}.}${}$
\label{fig:ECMA1}
\end{figure}

\citet{BuHaTiVo20} use 56 state clusters in the U.S.\footnote{The cluster variable is the state as in Section \ref{sec:states}. We selected this example, however, because the data are different and the numbers of state categories are different too.} for their analysis summarized in their Tables 1--2.
We extracted their cluster variable and draw its Hill plot for the largest 28 (50\%) clusters in Figure \ref{fig:ECMA1}.

For every number $k$ of the top order statistics except for $k=20$ and 21, the 95\% confidence interval does not exclude the possibility of $\beta < 2$.
Therefore, we cannot plausibly assume $\beta > 2$ for this data set to use the conventional CR methods.
Furthermore, using the formal statistical test presented in Appendix \ref{sec:test:appendix}, we reject the hypothesis of $\mathbb{E}[S_g^2]<\infty$ for a number of regressions whose results are reported in their Tables 1--2.

%%%%%%%%%%%%%%%%%%%%%%%%%%%%
\subsection{Region Clusters in Russia \citep{EnMaPe20}}
%%%%%%%%%%%%%%%%%%%%%%%%%%%%

\begin{figure}[h]
\vspace{1cm}\centering
\includegraphics[width=11cm]{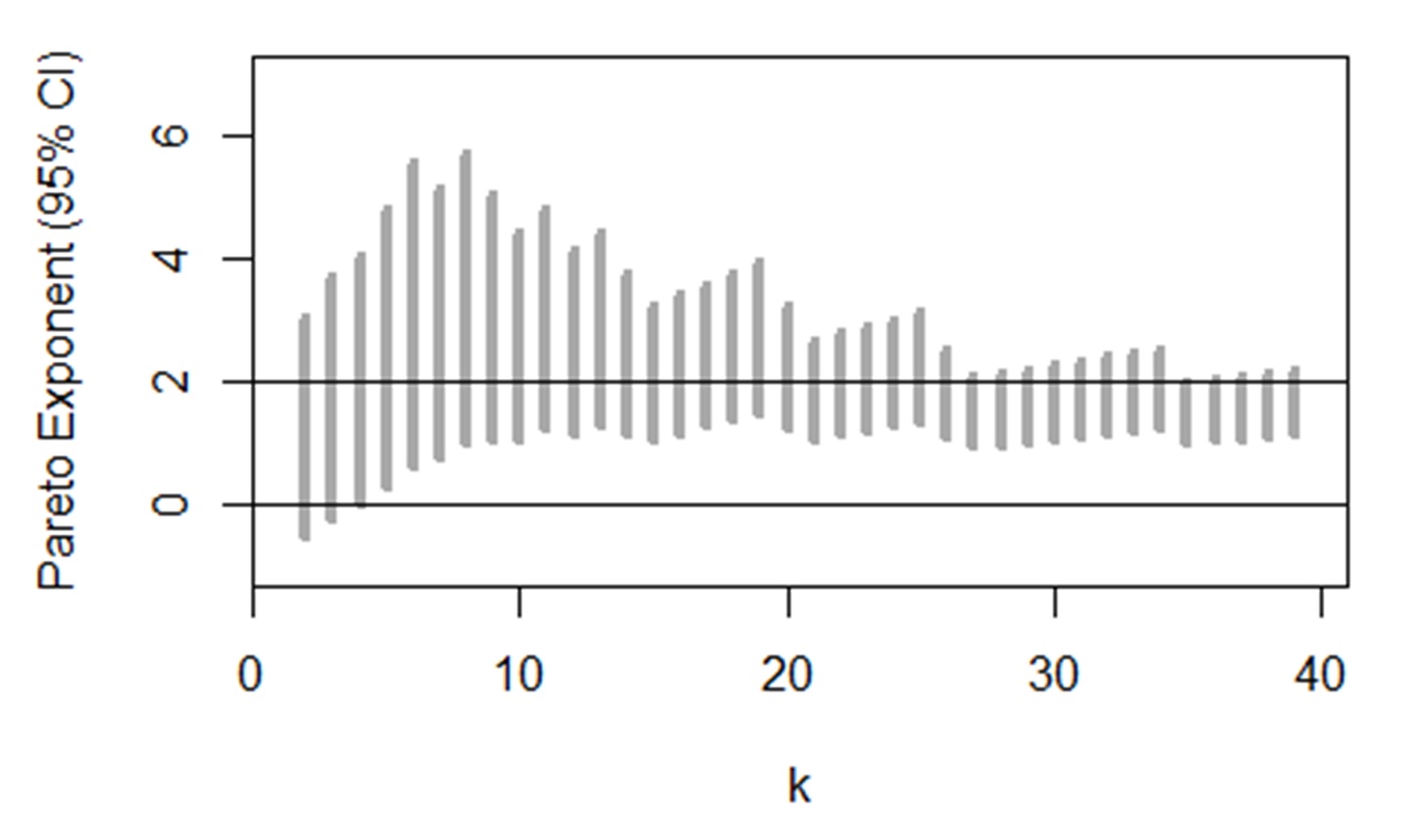}
\caption{The Hill plot for the largest 39 (50\%) of the 78 region clusters for \citep{EnMaPe20}.}${}$
\label{fig:ECMA2}
\end{figure}

\citet{EnMaPe20} use 78 region clusters in Russia for their analysis summarized in their Tables 1--3.
We extract their cluster variable and draw its Hill plot for the largest 39 (50\%) clusters in Figure \ref{fig:ECMA2}.

For every number $k$ of the top order statistics, the 95\% confidence interval does not exclude the possibility of $\beta < 2$.
Therefore, we cannot plausibly assume $\beta > 2$ for this data set to use the conventional CR methods.
Furthermore, using the formal statistical test presented in Appendix \ref{sec:test:appendix}, we reject the hypothesis of $\mathbb{E}[S_g^2]<\infty$ for a number of regressions whose results are reported in their Tables 1--3.

%%%%%%%%%%%%%%%%%%%%%%%%%%%%
\subsection{Branch Clusters \citep{AlBaBaBuRaSuVi20}}
%%%%%%%%%%%%%%%%%%%%%%%%%%%%

\begin{figure}[h]
\vspace{1cm}\centering
\includegraphics[width=11cm]{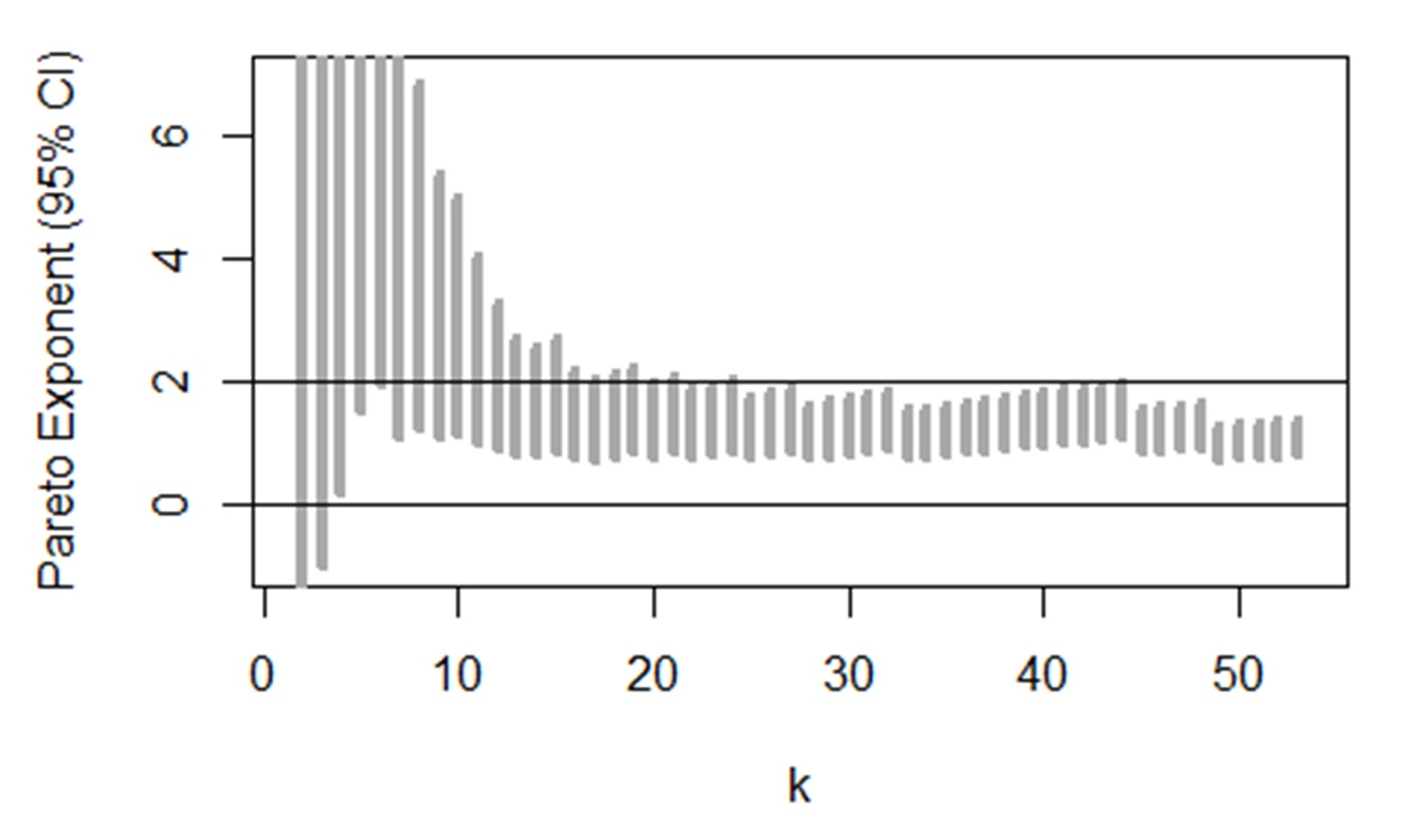}
\caption{The Hill plot for the largest 54 (50\%) of the 108 BRAC branch clusters for \citep{AlBaBaBuRaSuVi20}.}${}$
\label{fig:ECMA3}
\end{figure}

\citet{AlBaBaBuRaSuVi20} use 108 NGO (BRAC) branch clusters for for their analysis summarized in their Table 7.
We extract their cluster variable and draw its Hill plot for the largest 54 (50\%) clusters in Figure \ref{fig:ECMA3}.

For every number $k$ of the top order statistics, the 95\% confidence interval does not exclude the possibility of $\beta < 2$.
Therefore, we cannot plausibly assume $\beta > 2$ for this data set to use the conventional CR methods.
Furthermore, using the formal statistical test presented in Appendix \ref{sec:test:appendix}, we reject the hypothesis of $\mathbb{E}[S_g^2]<\infty$ for a number of regressions whose results are reported in their Table 7.

%%%%%%%%%%%%%%%%%%%%%%%%%%%%
\subsection{How Many Research Papers Cluster Standard Errors?}
%%%%%%%%%%%%%%%%%%%%%%%%%%%%

Once again, we want to stress that the above list of examples is just the tip of the iceberg.
Our selection of the three papers is due only to the relative ease of replication thanks to the efforts by the authors of these papers.
There were many other papers for which replication was difficult or infeasible, sometimes due to limited data availability.
It is worthy to note that there are 
at least
nine papers that cluster standard errors among those published in \textit{American Economic Review} in year 2020, and
at least
eight papers that cluster standard errors among those published in \textit{Econometrica} in year 2020, among others.\footnote{The results of the test of the null hypothesis $\mathbb{E}[S_g^2]<\infty$ for all these papers, as well as those in other top five journals, based on the method presented in Appendix \ref{sec:test:appendix}, will become available in spreadsheet upon request from the authors under certain conditions. We thank our graduate research assistants for their great efforts to replicate the analyses by the large number of papers.}

\section{Non-Gaussian Limit of Self-Normalized Means}\label{sec:self_normalized}
%%%%%%%%%%%%%%%%%%%%%%%%%%%%

When the score does not have a finite variance, one might still want to resort to the self-normalized central limit theorem (CLT) to establish the asymptotic normality under a possibly slower convergence rate of convergence than the standard $\sqrt{G}$ rate.
Under our framework in which the distribution of $\{N_g\}$ has a heavy tail, however, such a self-normalized CLT is not guaranteed to work.
It may or may not hold depending on the structure of within-cluster dependence.

For simplicity of illustration, suppose that one is interested in obtaining the standard error of the estimator $\hat\theta = N^{-1} \sum_{g=1}^G \sum_{i=1}^{N_g} Y_{gi}$ for the mean $\theta = \mathbb{E}[Y_{gi}]$.
In this setting, one may hope to establish the self-normalized asymptotic  normality $\mathbb{E}[(\hat\theta - \theta)^2]^{-1/2} (\hat\theta - \theta) \stackrel{d}{\rightarrow} \mathcal{N}(0,1)$, or more practically, $\widehat{\mathbb{E}}[(\hat\theta - \theta)^2]^{-1/2} (\hat\theta - \theta) \stackrel{d}{\rightarrow} \mathcal{N}(0,1)$.
Such a self-normalized CLT may fail to hold, however, if the distribution of $N_g$ has a heavy tail and within-cluster correlation is strong.

For instance, consider the data generating process
$
Y_{gi} = \rho_G R_g + e_{gi},
$
where $R_g$ is a cluster-specific effect which induces within-cluster dependence, and $e_{gi}$ is a random noise which is i.i.d. across $i$ and $g$.
In this case, the limit distribution of the self-normalized mean $\mathbb{E}[(\hat\theta - \theta)^2]^{-1/2} (\hat\theta - \theta)$ in general fails to be asymtotically Gaussian if $\rho_G \gg G^{1/2-1/\beta} \rightarrow 0$ as $G\rightarrow\infty$ when $\beta<2$.
See Appendix \ref{sec:appendix:self_normalized} for details about this argument.
Hence, the rate-adaptive CR standard errors and even the jackknife standard errors may fail to work in our framework of cluster sampling involving cluster-specific effects and heavy-tailed distributions of cluster sizes.

Recall that cluster-specific variables were the initial motivations for applied researchers to start using cluster-robust standard errors in cross-section data \citep[e.g.,][]{He1998}.
Cluster-specific variables are ubiquitous in economic data set.
This is also true by design for cluster randomized experiments.

%%%%%%%%%%%%%%%%%%%%%%%%%%%%
\section{Our Proposal of a New Robust Approach}\label{sec:wcr}
%%%%%%%%%%%%%%%%%%%%%%%%%%%%
This section proposes a new approach to cluster-robust inference.
It does not suffer from the aforementioned problem with the standard CR methods characterized by Theorem \ref{prop:tail}.
Our proposal is simple, but we provide a complete theoretical rationale for why it works unlike the standard CR approach.

One simple way to fix the problem of the non-robustness of the standard CR methods is to modify \eqref{eq:ols} and \eqref{eq:cr} by
\begin{align*}
&\widehat\theta^{\text{WCR}} = \left(\sum_{g=1}^G N_g^{-1} X_g'X_g\right)^{-1} \left(\sum_{g=1}^G N_g^{-1} X_g'Y_g\right)
\qquad\text{and}\\
&\widehat V_{\hat\theta}^{\text{WCR}} = a_n \left(\sum_{g=1}^G N_g^{-1} X_g'X_g\right)^{-1} \left(\sum_{g=1}^G N_g^{-2} \widehat S_g \widehat S_g' \right) \left(\sum_{g=1}^G N_g^{-1} X_g'X_g\right)^{-1},
\end{align*}
respectively, where $a_n \overset{p}\rightarrow 1$.

Likewise, we may also modify the jackknife estimator \eqref{eq:crjack} by
\begin{align*}
\widehat V_{\hat\theta}^{\text{WCR,JACK}} = \sum_{g=1}^G \left(\widehat\theta_{-g}^{\text{WCR}}-\widehat\theta^{\text{WCR}}\right) \left(\widehat\theta_{-g}^{\text{WCR}}-\widehat\theta^{\text{WCR}}\right)',
\end{align*}
where 
$$
\widehat\theta_{-g}^{\text{WCR}} = \left(\sum_{h \neq g} N_h^{-1}X_h'X_h\right)^{-1} \left(\sum_{h \neq g} N_h^{-1}X_h'Y_h\right)
$$
denotes the weighted leave-one-cluster-out estimator.

We are not the first to suggest a weighted estimator.
In the different context of cluster randomized experiments, \citet*{BuCaShTa2022} suggest a weighted difference-in-mean estimator similarly to ours.
While they propose such a weighting for the purpose of identifying and estimating certain parameter of interest, our estimator $\widehat\theta^{\text{WCR}}$ in fact encompasses their estimator $\hat\theta_{1,G}$ as a special case.\footnote{They propose two kinds of estimators. While their $\hat\theta_{1,G}$ is analogous to our estimator $\widehat\theta^{\text{WCR}}$, their $\hat\theta_{2,G}$ is analogous to the conventional CR estimator $\widehat\theta$.}
This implies that their estimator works under our assumptions as well. 
See Remark \ref{remark:BCST} ahead.

As in Section \ref{sec:pitfall}, for given $j \in \{1,\cdots,\dim\{X_{gi}\}\}$, let $\Sigma_g$ and $Z_{gi}$ be short-hand notations for the $j$-the coordinate of $S_g$ and the $j$-th coordinate of $X_{gi}U_{gi}$, respectively.
Define $\widetilde \Sigma_g := \Sigma_g/N_g = \sum_{i=1}^{N_g} Z_{gi}$. 
In contrast to Theorem \ref{prop:tail}, the following theorem establishes that $\widetilde \Sigma_g$ has the same tail index as the original score $Z_{gi}$. 

\begin{theorem}
\label{prop:tail2} 
If Assumption \ref{a:cluster} holds, then for any $z>0$, 
\[
\frac{\mathbb{P}(\widetilde \Sigma_g > zt)}{\mathbb{P}(\widetilde \Sigma_g > t)} = z^{-\alpha} \text{ as } t\rightarrow \infty.
\]
\end{theorem}

\noindent
We present a proof in Appendix \ref{sec:prop:tail2}. 

\begin{remark}[Relaxing the Assumption of Independent Cluster Sizes]\label{remark:relaxing}
Recall that Assumption \ref{a:cluster}.\ref{a:cluster:indep} requires that $N_g$ is independent of $(Z_{g1},Z_{g2},\ldots)$.
This assumption may be too strong in certain applications.
\citet*{BuCaShTa2022} for instance consider non-independent cluster sizes for cluster randomized experiments.
In the statement of Theorem \ref{prop:tail2}, Assumption \ref{a:cluster}.\ref{a:cluster:indep} may be relaxed by strengthening Assumption \ref{a:cluster}.\ref{a:cluster:copula}.
Namely, suppose that the conditional copula density of $(Z_{g1},Z_{g2},\ldots)$ given $N_g$ exists and is uniformly bounded, $\mathbb{E}[|Z_{ig}| \ |N_g]<\infty$ almost surely, and the tail index of the conditional distribution of $Z_{ig}$ given $N_g$ is $\alpha$ for all $N_g$, then we can remove the independence requirement from Assumption \ref{a:cluster}.\ref{a:cluster:indep}.
\end{remark}

Recall that the non-robust methods (e.g., those based on Eicker-Huber-White standard errors) requires only bounded second moments of the score $X_{gi} U_{gi}$.
Also recall from Section \ref{sec:pitfall} that the conventional CR methods can fail when the second moment of $N_g$ is infinite \textit{even if} the second moments of the score $X_{gi} U_{gi}$ were finite.
Now, Theorem \ref{prop:tail2} states that our proposed robust approach works as far as the second moments of the score $X_{gi} U_{gi}$ is bounded regardless of whether the second moment of $N_g$ if finite or not. 
Accordingly, we allow $\sup_g N_g^2/N$ to diverge. 

To derive the asymptotic properties of our WCR estimator, we make the following additional assumption. 

\begin{assumption}\label{a:normal}
1. $(N_g, X_g, Y_g)$ is i.i.d. across $g$.
2. $\mathbb{E}[N_g^{-1} X_g'X_g]$ is non-singular.
\end{assumption}

Part 1 of this assumption requires the i.i.d. sampling \textit{across} clusters, as is assumed in general under cluster sampling environments.
It does, however, allow for arbitrary dependence \textit{within} each cluster as in the general cluster sampling environments.
Part 2 of the above assumption rules out multi-collinearity.
In summary, Assumption \ref{a:normal} is a standard requirement.

Recall that our main claim from Section \ref{sec:pitfall} is that the asymptotic normality \eqref{eq:asymptotic} for the conventional CR inference may not work if $\beta < 2$ \textit{even if} $\alpha > 2$.
Section \ref{sec:states} demonstrates that we cannot rule out such a pathetic case with $\beta < 2$ even for the most common setting of cluster sampling, namely using 51 states as clusters.
Furthermore, Section \ref{sec:self_normalized} argues that even the self-normalized CLT generally fails for the conventional CR inference.
On the other hand, the following theorem shows that our proposed CR approach based on $\widehat\theta^{\text{WCR}}$ and $\widehat{V}_{\hat\theta}^{\text{WCR}}$ works as far as $\alpha > 2$ is true, regardless of whether $\beta < 2$ is true or not.

\begin{theorem}
\label{prop:wcr}
If Assumptions \ref{a:cluster} and \ref{a:normal} hold with $\alpha > 2$, then
\begin{equation*}
\sqrt{G}(\widehat\theta^{\text{WCR}} - \theta) \xrightarrow{p} \mathcal{N}(0,V^{\text{WCR}})
\end{equation*}
as $G\rightarrow\infty$,
where
\begin{equation*}
V^{\text{WCR}} =  (\mathbb{E}[N_g^{-1} X_g'X_g])^{-1}(\mathbb{E}[N_g^{-2} X_g 'X_gU^2_g])(\mathbb{E}[N_g^{-1} X_g'X_g])^{-1}.
\end{equation*}
Furthermore, we have
$
G\widehat V_{\hat\theta}^{\text{WCR}} \xrightarrow{p} V^{\text{WCR}}
$
 as $G\rightarrow\infty$
\end{theorem}  

\noindent
A proof is presented in Appendix \ref{sec:prop:wcr}.
A few remarks are in order.

\begin{remark}[Relaxing the Assumption of Independent Cluster Sizes]\label{remark:relaxing2}
Remark \ref{remark:relaxing} applies to Theorem \ref{prop:wcr} too.
Namely, we can relax the requirement of independent cluster sizes in Assumption \ref{a:cluster}.\ref{a:cluster:indep} at the cost of strengthening Assumption \ref{a:cluster}.\ref{a:cluster:copula}.
\end{remark}

\begin{remark}[Relaxing the Assumption on Individual Scores]
The conditions of $\alpha > 2$, as well as $\beta > 2$, may be relaxed by considering a self-normalized version of Theorem \ref{prop:wcr}, although the $\sqrt{G}$ rate of convergence would be generally lost in that case.
\end{remark}

\begin{remark}[Thin-Tailed Distributions of Cluster Sizes]
The RV condition in Assumption \ref{a:cluster}.\ref{a:cluster:rv} is assumed only for concise statement and is not essential.
We can more trivially accommodate thin-tailed distributions, such as the Gaussian distribution, for the individual score.
\end{remark}

\begin{remark}\label{remark:BCST}
As mentioned earlier, our estimator $\widehat\theta^{\text{WCR}}$ encompasses the estimator $\hat\theta_{1,G}$ of \citet*{BuCaShTa2022} as a special case.
While \citet*{BuCaShTa2022} assume that $\sup_g N_g^2/N \stackrel{p}{\rightarrow} 0$, our Theorem \ref{prop:wcr} implies that their estimator $\hat\theta_{1,G}$ in fact works even under heavy-tailed distributions of $N_g$.
In other words, our result shows that their estimator $\hat\theta_{1,G}$ is in fact robust to heavy-tailed distribution of $N_g$.
\end{remark}

As a consequence of Theorem \ref{prop:wcr}, one can use
$$
\widehat\theta^{\text{WCR}}_j \pm 1.96 \sqrt{\widehat V_{\hat\theta, jj}^{\text{WCR}}}
\qquad\text{ or }\qquad
\widehat\theta^{\text{WCR}}_j \pm 1.96 \sqrt{\widehat V_{\hat\theta, jj}^{\text{WCR,JACK}}}
$$
as a robust 95\% confidence interval for the $j$-th coordinate of $\theta$ even if $\beta < 2$ as is likely the case with 51 states used as clusters.

To justify the use of the jackknife WCR variance estimator $\widehat V_{\hat\theta, kk}^{\text{WCR,JACK}}$, we close this section with the following proposition.

\begin{proposition}\label{prop:jackknife}
If Assumptions \ref{a:cluster} and \ref{a:normal} hold with $\alpha > 2$, then 
$
G\widehat{V}_{\hat{\theta}}^{\text{WCR,JACK}} \overset{p}{\rightarrow }V^{\text{WCR}}
$
as $G \rightarrow \infty$.
\end{proposition}  

\noindent
A proof is presented in Appendix \ref{sec:prop:jackknife}.

%%%%%%%%%%%%%%%%%%%%%%%%%%%%
\section{Simulations}\label{sec:simulation}
%%%%%%%%%%%%%%%%%%%%%%%%%%%%

In this section, we present simulation studies to evaluate the finite sample performance of our proposed WCR methods in comparison with the conventional CR methods.
In light of the discussions in Section \ref{sec:self_normalized}, we focus on data that contain cluster-specific variables.

The data generating design is defined as follows.
We consider the cluster treatment model with individual covariates
$$
Y_{gi} = \theta_0 + \theta_1 T_g + \sum_{j=1}^K \theta_j X_{g,j+1} + U_{gi}
$$
following \citet*[][Equation (40)]{MaNiWe2022fast} among others.
The binary treatment variable $T_g$ takes the value of one for $\lceil 0.2G \rceil$ clusters and zero for the remaining clusters $G - \lceil 0.2G \rceil$, where $\lceil a \rceil$ denotes the smallest integer greater than or equal to $a$.
We draw cluster sizes 
$
N_g \sim \lceil 10 \cdot \text{Pareto}(1,\beta) \rceil
$ 
independently for $g \in \{1,\cdots,G\}$.
For each $g \in \{1,\cdots,G\}$, we independently draw $N_g$-variate random vectors,
$
(\tilde X_{g1j},\cdots, \tilde X_{gN_gj})' \sim \mathcal{N}(0,\Omega)
$
for $j \in \{1,\cdots,K\}$ and
$
(\tilde U_{g1},\cdots,\tilde U_{gN_g})' \sim \mathcal{N}(0,\Omega),
$
where $\Omega$ is an $N_g \times N_g$ variance-covariance matrix such that $\Omega_{ii}=1$ for all $i \in \{1,\cdots,N_g\}$ and $\Omega_{ij}=1/2$ whenever $i \neq j$.
The controls are constructed by
$
X_{gik} = 0.2 F_{\text{Beta}(2,2)}^{-1} \circ \Phi (\tilde X_{gik}),
$
where $F_{\text{Beta}(2,2)}$ and $\Phi$ denote the CDFs of the $\text{Beta}(2,2)$ and standard normal distributions, respectively.
The errors are heteroskedastically constructed by
$
U_{gi} = 0.2 \tilde U_{gi}
$
if $T_g=0$
and
$
U_{gi} = \tilde U_{gi}
$
if $T_g=1$.

We vary values of the exponent parameter $\beta \in \{1,2,4\}$ across sets of simulations.
The regression coefficients are fixed to $(\theta_0,\theta_1,\theta_2,\cdots,\theta_{K+1})'=(1,1,1,\cdots,1)'$ throughout, whereas the dimension $K$ of covariates vary as $K \in \{0,1,5\}$.
We set the sample size (i.e., the number of clusters) to $G = 50$ across sets of simulations, which is close to the number of states in the U.S. we discussed as an example earlier.
Each set of simulations consists of 10,000 Monte Carlo iterations.

%%%%%%%%%%%%%%%%%%%%%%%%%%%%%%%%%%%%%%%%%%%%%%%%%%%
\begin{figure}
\bigskip\centering
$\beta=2$\\
\includegraphics[width=13.25cm]{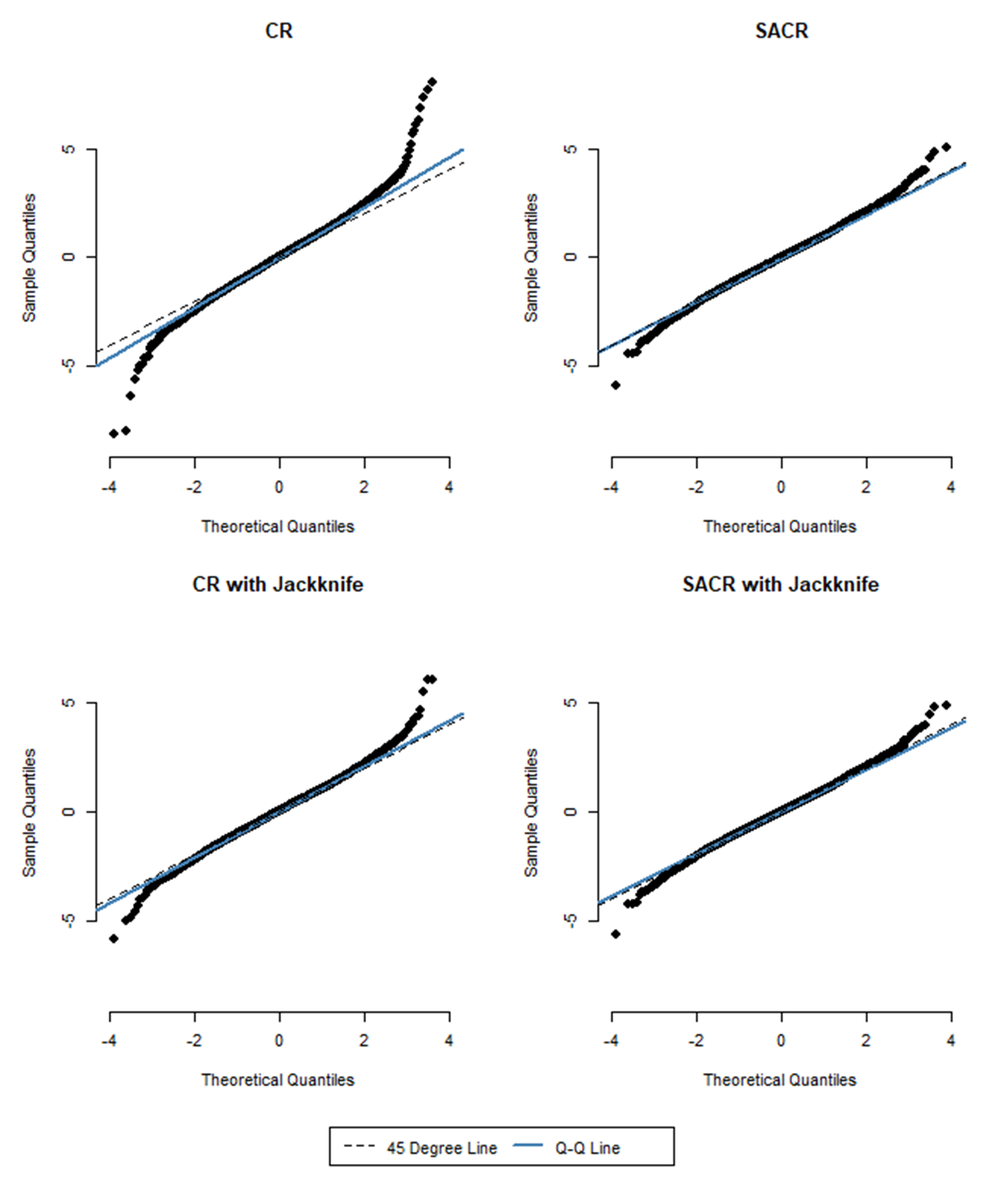}
\caption{Q-Q plots for the self-normalized statistics under $\beta=2$. The dashed line indicates the $45^{\circ}$ line, and the solid line indicates the fitted line. The left column shows the results for the conventional CR methods, while the right column shows the results for our proposed WCR methods. The top row shows the results with analytic standard errors, while the bottom row shows the results with jackknife estimators.}${}$
\label{fig:QQ2.0}
\end{figure}
%%%%%%%%%%%%%%%%%%%%%%%%%%%%%%%%%%%%%%%%%%%%%%%%%%%
\begin{figure}
\bigskip\centering
$\beta=1$\\
\includegraphics[width=13.25cm]{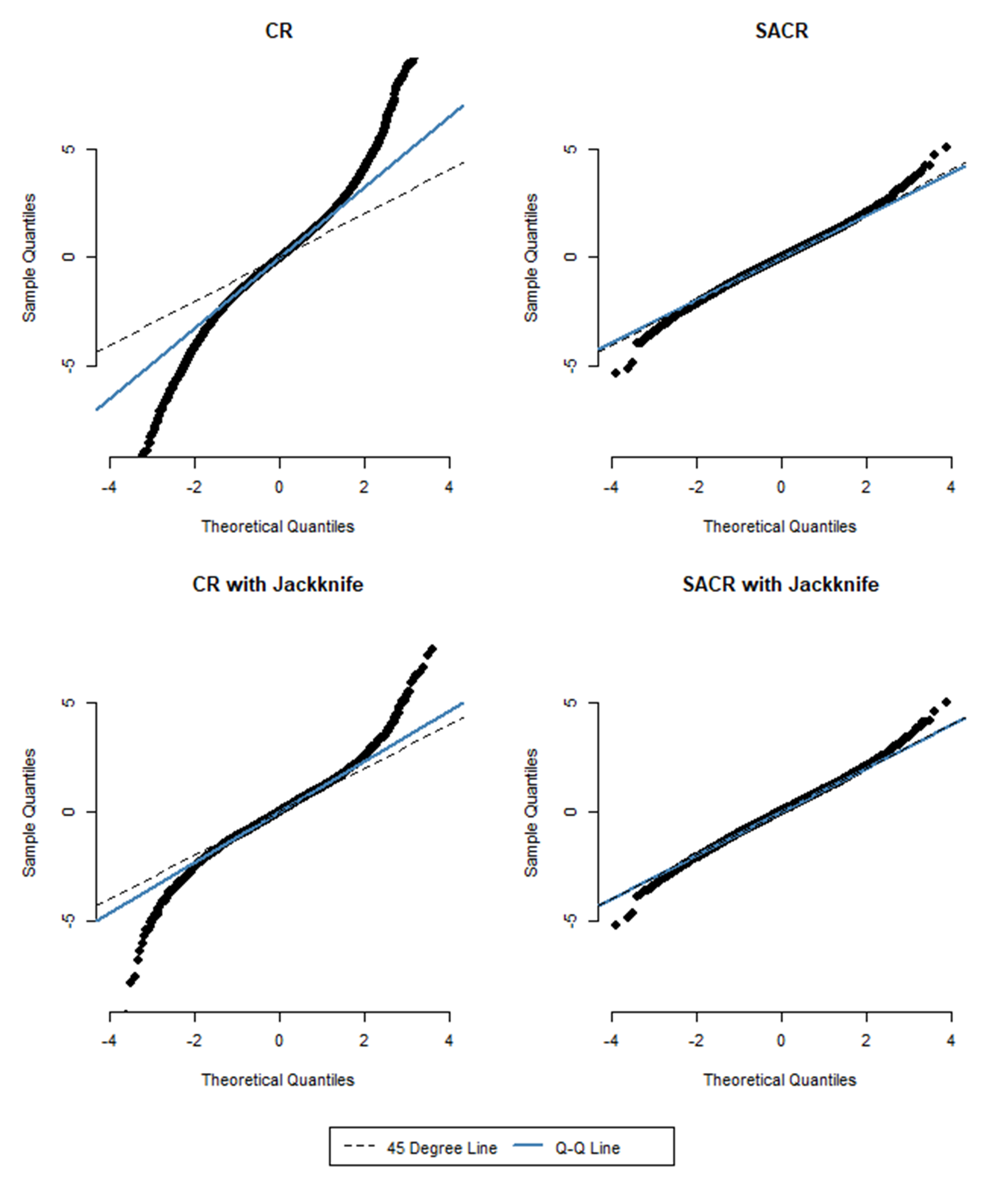}
\caption{Q-Q plots for the self-normalized statistics under $\beta=1$. The dashed line indicates the $45^{\circ}$ line, and the solid line indicates the fitted line. The left column shows the results for the conventional CR methods, while the right column shows the results for our proposed WCR methods. The top row shows the results with analytic standard errors, while the bottom row shows the results with jackknife estimators.}${}$
\label{fig:QQ1.0}
\end{figure}
%%%%%%%%%%%%%%%%%%%%%%%%%%%%%%%%%%%%%%%%%%%%%%%%%%%

Figures \ref{fig:QQ2.0}--\ref{fig:QQ1.0} draw Q-Q plots under $\beta=2$ and $\beta=1$, respectively, of the self-normalized statistics:
\begin{align*}
\text{(CR)}\qquad
&\left.\left( \widehat\theta_1 - \theta_1 \right) \right/ \sqrt{\widehat V_{\hat\theta,11}^{\text{CR}}},
\\
\text{(CR Jackknife)}\qquad
&\left.\left( \widehat\theta_1 - \theta_1 \right) \right/ \sqrt{\widehat V_{\hat\theta,11}^{\text{CR,JACK}}},
\\
\text{(WCR)}\qquad
&\left.\left( \widehat\theta_1^{\text{WCR}} - \theta_1 \right) \right/ \sqrt{\widehat V_{\hat\theta,11}^{\text{WCR}}}, \qquad\text{and}
\\
\text{(WCR Jackknife)}\qquad
&\left.\left( \widehat\theta_1^{\text{WCR}} - \theta_1 \right) \right/ \sqrt{\widehat V_{\hat\theta,11}^{\text{WCR,JACK}}}.
\end{align*}
For these figures, we focus on the case with $K=0$.
In each figure, the dashed line indicates the $45^{\circ}$ line, and the solid line indicates the fitted line.

Observe that the self-normalized statistics based on the conventional CR methods suffer from farther deviation away from the theoretical quantiles, whereas those based on our proposed WCR methods more precisely follow the theoretical quantiles.
This observation is true for both the analytic standard error estimator and the jackknife estimator.
The deviations for the conventional CR methods further exacerbate as the distribution of $N_g$ becomes heavier, as in the transition from $\beta=2$ (in Figure \ref{fig:QQ2.0}) to $\beta=1$ (in Figure \ref{fig:QQ1.0}).
These results are consistent with the general non-Gaussianity of the conventional CR methods as discussed in Section \ref{sec:self_normalized}, as well as the guaranteed Gaussianity of our proposed WCR methods as discussed in Section \ref{sec:wcr}.

Table \ref{tab:sim} summarizes simulation results. 
Displayed are 
the mean square error (MSE), 
the rejection frequencies based on the analytic standard error estimation in round brackets, and 
the rejection frequencies based on the jackknife standard error estimation in square brackets.
The nominal probability of rejection is set to $p=0.05$ throughout.
The first three columns show the results for the conventional CR methods.
The last three columns show the results for our proposed WCR methods.

%%%%%%%%%%%%%%%%%%%%%%%%%%%%%%%%%%%%%%%%%%%%%%%%%%%
\begin{table}
\centering
\renewcommand{\arraystretch}{1.4}
\begin{tabular}{cccccccccc}
\hline\hline
    &     &         & \multicolumn{3}{c}{Conventional CR} && \multicolumn{3}{c}{New WCR}\\
\cline{4-6}\cline{8-10}
    &     &         & ${}$ \ \ MSE \ \ ${}$ & \multicolumn{2}{c}{Rejection ({\small $p=0.050$})} && ${}$ \ \ MSE \ \ ${}$ & \multicolumn{2}{c}{Rejection ({\small $p=0.050$})}\\
\cline{5-6}\cline{9-10}
$K$ & $\beta$ && $\widehat\theta_1$ & $\widehat V_{\hat\theta,11}^{\text{CR}}$ & $\widehat V_{\hat\theta,11}^{\text{CR,JACK}}$ && $\widehat\theta_1^{\text{WCR}}$ & $\widehat V_{\hat\theta,11}^{\text{WCR}}$ & $\widehat V_{\hat\theta,11}^{\text{WCR,JACK}}$\\
\hline
  0 & 4 && 0.057 & (0.095) & [0.072] && 0.054 & (0.088) & [0.067]\\
    & 2 && 0.077 & (0.141) & [0.088] && 0.055 & (0.086) & [0.069]\\
    & 1 && 0.144 & (0.272) & [0.106] && 0.053 & (0.073) & [0.068]\\
\hline
  1 & 4 && 0.058 & (0.096) & [0.073] && 0.054 & (0.088) & [0.068]\\
    & 2 && 0.074 & (0.136) & [0.085] && 0.054 & (0.087) & [0.070]\\
    & 1 && 0.138 & (0.273) & [0.108] && 0.053 & (0.074) & [0.070]\\
\hline
  5 & 4 && 0.057 & (0.094) & [0.065] && 0.054 & (0.082) & [0.063]\\ 
    & 2 && 0.071 & (0.130) & [0.082] && 0.053 & (0.079) & [0.064]\\
    & 1 && 0.121 & (0.254) & [0.101] && 0.053 & (0.070) & [0.068]\\
%\hline
% 10 & 4 && 0.059 & (0.098) & [0.074] && 0.056 & (0.091) & [0.073]\\ 
%    & 2 && 0.068 & (0.125) & [0.078] && 0.055 & (0.082) & [0.069]\\
%    & 1 && 0.110 & (0.235) & [0.099] && 0.053 & (0.066) & [0.070]\\
\hline\hline
\end{tabular}
\caption{\setlength{\baselineskip}{6.5mm}Simulation results based on 10,000 Monte Carlo iterations. Displayed are the mean square error (MSE), the rejection frequencies based on the analytic standard error estimation in round brackets  with the nominal probability of $0.050$, and the rejection frequencies based on the jackknife standard error estimation in square brackets with the nominal probability of $0.050$. The first three columns show the results for the conventional CR methods, and the last three columns show the results for our proposed WCR method.}${}$
\label{tab:sim}
\end{table}
%%%%%%%%%%%%%%%%%%%%%%%%%%%%%%%%%%%%%%%%%%%%%%%%%%%

We find the following three observations in these simulation results.
First, focus on the MSE.
While the MSE of the OLS estimator $\widehat\theta_1$  exponentially blows up as $\beta$ decreases,
the MSE of the WCR estimator $\widehat\theta_1^{\text{WCR}}$ remains stable as $\beta$ varies.
Second, consider the rejection frequencies reported in the round brackets based on the analytic standard error estimators, $\widehat V_{\hat\theta,11}^{\text{CR}}$ and $\widehat V_{\hat\theta,11}^{\text{WCR}}$.
While the rejection frequencies for the conventional CR method based on $\widehat V_{\hat\theta,11}^{\text{CR}}$ blows up as $\beta$ decreases,
the rejection frequencies for our proposed WCR method based on $\widehat V_{\hat\theta,11}^{\text{WCR}}$ remain stable and closer to the nominal rejection probability of $p=0.050$ as $\beta$ varies.
Third, consider the rejection frequencies reported in the square brackets based on the jackknife standard error estimators, $\widehat V_{\hat\theta,11}^{\text{CR,JACK}}$ and $\widehat V_{\hat\theta,11}^{\text{WCR,JACK}}$.
While these jackknife standard error estimators deliver more desirable rejection frequencies than the analytic standard error estimators for each of the conventional CR method and the new WCR method, we continue to observe the same qualitative pattern as in the case of the analytic estimators.
Namely, while the rejection frequencies for the conventional CR method based on $\widehat V_{\hat\theta,11}^{\text{CR,JACK}}$ blows up as $\beta$ decreases,
the rejection frequencies for our proposed WCR method based on $\widehat V_{\hat\theta,11}^{\text{WCR,JACK}}$ remain stable and closer to the nominal rejection probability of $p=0.050$ as $\beta$ varies.

From these observations, it seems more desired to use the new WCR methods over the conventional CR methods for estimation accuracy as well as the property of robust inference.

%%%%%%%%%%%%%%%%%%%%%%%%%%%%
\section{Summary and Discussions}
%%%%%%%%%%%%%%%%%%%%%%%%%%%%

CR standard errors are used extensively in the practice of empirical economic analyses.
Despite its wide practice, the literature has been silent about a potential pitfall of the conventional CR methods.
The theory for the conventional CR methods assumes that (i) $N_g$ is fixed while $G \rightarrow \infty$, or at best (ii) $\sup_g N_g^2/N \rightarrow 0$ as $G \rightarrow \infty$.
Even the most common empirical settings, such as the case of using the 51 states in the U.S. as clusters, violate these existing assumptions.

We establish three theoretical results in this paper.
First, if the distribution of $N_g$ follows the power low with exponent less than two, then the second moment of the CR score fails to exist \textit{even if} the second moment of the individual score were finite.
Second, the second moment of our proposed WCR score exists as far as the second moment of the individual score exists \textit{regardless of} the exponent of the distribution of $N_g$.
Third, consequently, our proposed WCR methods enjoy robust inference unlike the conventional CR methods.
It is also worthy to note that, in the special case where the tail exponent $\beta$ is large, the WCR methods behave similarly to the conventional CR methods.

One may want to resort to self-normalized CLT to validate the conventional CR standard errors and the jackknife estimators. 
However, the limit distribution of self-normalized sums is not guaranteed to be Gaussian under the power law of cluster sizes.
Hence, we cannot recommend using the conventional CR methods if one is interested in robustness in inference under a small number of large clusters, such as the case of using the 51 states in the U.S. as clusters.
A main disadvantage of our proposed methods is that the weighting alters the estimator from our familiar OLS estimator, which is a cost that a researcher needs to pay to enjoy the robustness.

\vspace{0cm}
%%%%%%%%%%%%%%%%%%%%%%%%%
\setlength{\baselineskip}{7.5mm}
\bibliographystyle{ecta}
\bibliography{bib}
\setlength{\baselineskip}{7.9mm}
%%%%%%%%%%%%%%%%%%%%%%%%%
\vspace{0cm}

%%%%%%%%%%%%%%%%%%%%%%%%%%%%
\appendix
\section*{Appendix}
%%%%%%%%%%%%%%%%%%%%%%%%%%%%

This appendix section presents omitted details.
Appendix \ref{sec:proof} presents mathematical proofs.
Appendix \ref{sec:test:appendix} presents a test of robustness.
Appendix \ref{sec:appendix:self_normalized} presents details of Section \ref{sec:self_normalized}.
Throughout, we use the following short-hand notations.
$a=o_p(1)$ means that $a$ converges in probability to zero,
$a=O_p(1)$ means that $a$ is stochastically bounded, and
$a \simeq_p b$ means that $a/b=O_p(1)$ and $b/a=O_p(1)$.

%%%%%%%%%%%%%%%%%%%%%%%%%%%%
\section{Proofs}\label{sec:proof}
%%%%%%%%%%%%%%%%%%%%%%%%%%%%
\subsection{Proof of Theorem \ref{prop:tail}}\label{sec:prop:tail}
%%%%%%%%%%%%%%%%%%%%%%%%%%%%
\begin{proof}
For the first part, since $\beta>1$, we have that $\mathbb{E}[N_g]<\infty$ and it follows from the strong law of large numbers that $N/G=G^{-1}\sum^G_{g=1}N_g \rightarrow \mathbb{E}[N_g]$ almost surely. 
Given the regularly varying tail, we have that $\sup_g N_g \simeq_p G^{1/\beta}$, yielding that $\sup_g N^2_g/N \simeq_p G^{2/\beta-1}$, which diverges with probability approaching one since $\beta<2$. 

For the second part, we say that sequence $\{Z_{gi}\}_{i=1}^{n}$ of random variables is extended negatively dependent (END) if there exists a finite constant $M>0$ such that both
\[
\mathbb{P}\left( Z_{g1}\leq z_{1},...,Z_{gn}\leq z_{n}\right) \leq M\prod_{i=1}^{n}\mathbb{P}\left( Z_{gi}\leq z_{i}\right)
\]%
and%
\[
\mathbb{P}\left( Z_{g1}>z_{1},...,Z_{gn}>z_{n}\right) \leq M\prod_{i=1}^{n} \mathbb{P}\left( Z_{gi}>z_{i}\right)
\]%
hold for all $z_{1},...,z_{n}$.

Assumption \ref{a:cluster}.\ref{a:cluster:rv} implies $\mathbb{E}\left[ N_g\right] <\infty $. 
Assumption \ref{a:cluster}.\ref{a:cluster:copula} implies that $Z_{g1},\dots ,Z_{gn}$ are END for every $n\geq 1$ \citep[][Example 4.2]{Liu09}. 
Finally, Assumption \ref{a:cluster}.\ref{a:cluster:indep} and $\beta < \alpha$ imply $\overline{F}\left( x\right) = 1 - F(x) = o\left( \overline{H}\left(x\right) \right) $. 
Then, \citet[][Theorem 3.2]{Ch15} yields that
\begin{equation}\label{eq:proof1}
\mathbb{P}(\Sigma_g >x)\sim \overline{H}\left( x/\mathbb{E}\left[ N_g\right] \right) \sim \overline{H}\left( x\right) \text{ as } x\rightarrow\infty,
\end{equation}
where $A\sim B$ means that $\lim A/B\rightarrow 1$ as $x \rightarrow \infty$. 
Note also that Assumption \ref{a:cluster}.\ref{a:cluster:indep} implies that for any $x>0$,
\begin{equation}\label{eq:proof2}
\frac{\overline{H}\left( xt\right) }{\overline{H}\left( t\right) }=x^{-\beta}\text{ as }t\rightarrow \infty.
\end{equation}
The proof is then complete by combining \eqref{eq:proof1} and \eqref{eq:proof2}.
\end{proof}

\subsection{Proof of Theorem \ref{prop:tail2}}\label{sec:prop:tail2}
%%%%%%%%%%%%%%%%%%%%%%%%%%%%
\begin{proof}
Throughout the proof, we let $C$ denote a generic finite constant, whose value could change line-by-line. 
Define 
\begin{equation*}
\theta_i= \frac{1}{N_g}\mathbf{1}[i \leq N_g]
\end{equation*}
for each $i \in \mathbb{N}$, 
where $\mathbf{1}[\cdot]$ denotes the indicator function. 
Then, we can write
\begin{equation*}
\widetilde \Sigma_g = \sum_{i=1}^{N} \frac{Z_{gi}}{N_g} = \sum_{i=1}^{\infty}\theta_{i} Z_{gi}. 
\end{equation*}

We first show that 
\begin{equation}
\label{eq:theta_mom1}
\sum_{i=1}^{\infty}\left(\mathbb{E}[\theta_i^{\alpha-\epsilon}]\right)^{\frac{1}{\alpha+\epsilon}}<\infty
\end{equation}
holds for some $\epsilon \in [0,\alpha]$.
To this end, we use Assumption \ref{a:cluster}.\ref{a:cluster:indep} to obtain 
\begin{align*}
\mathbb{E}\left[\theta_i^{\alpha-\epsilon}\right] 
&=\, \mathbb{E}\left[\frac{1}{N_g^{\alpha-\epsilon}}\mathbf{1}[i \leq N_g]\right] \\
&\leq\, \mathbb{E}\left[\frac{1}{i^{\alpha-\epsilon}}\mathbf{1}[i \leq N_g]\right] \\
&=\, \frac{1}{i^{\alpha-\epsilon}}\mathbb{P}[i \leq N_g] \\
&\, \leq C i^{-\beta-\alpha+\epsilon}.
\end{align*}
Thus, the inequality \eqref{eq:theta_mom1} follows by $\beta > \alpha > 1$. 

Second, we show that
\begin{equation}
\label{eq:theta_mom2}
\sum_{i=1}^{\infty} \mathbb{P}\left(\theta_i Z_{gi} >z \right)<\infty
\end{equation}
holds for any $z$.
Using Markov's inequality and the fact that $\mathbb{E}[Z_{gi}]<\infty$ holds by Assumption \ref{a:cluster}.\ref{a:cluster:rv}, we obtain
\begin{align*}
\sum_{i=1}^{\infty} \mathbb{P}\left(\theta_i Z_{gi} >z \right) 
\leq &\, \sum_{i=1}^{\infty}  \frac{\mathbb{E}\left[N_g^{-1}\mathbf{1}[i\leq N_g] Z_{gi} \right]}{z} \\
\leq &\, \sum_{i=1}^{\infty}  \frac{ C\mathbb{E}[Z_{gi}] i^{-1-\beta}}{z} \\
\leq &\, C < \infty. 
\end{align*}

Finally, the extended negatively dependence condition is sufficient for the tail independence as required by equation (2.9) in \citet{ZhShWe09}. 
By our equations \eqref{eq:theta_mom1} and \eqref{eq:theta_mom2}, their Conditions H1 and H2 are satisfied. 
Then, their Theorem 3.1 and Remark 3.2 yield
\begin{equation*}
\mathbb{P}\left( \widetilde \Sigma_g > z \right) \sim (1-F(z)) \sum_{i=1}^{\infty} \mathbb{E}[\theta_i^\alpha] \text{ as } z\rightarrow \infty,
\end{equation*}
where $\sum_{i=1}^{\infty} \mathbb{E}[\theta_i^\alpha]$ is finite by similar lines of argument to those used to derive \eqref{eq:theta_mom1} above. 
The proof is now complete by Assumption \ref{a:cluster}.\ref{a:cluster:indep}. 
\end{proof}

\subsection{Proof of Theorem \ref{prop:wcr}}\label{sec:prop:wcr}
%%%%%%%%%%%%%%%%%%%%%%%%%%%%
\begin{proof}
We first establish the asymptotic normality of $\widehat\theta^{\text{WCR}}$. 
Given Assumption \ref{a:normal}, it suffices to establish
\begin{equation}
\mathbb{E}[N_g^{-1} X'_g X_g] < \infty \label{eq:momentXX} 
\end{equation}
\begin{equation}
\mathbb{E}[N_g^{-2} X'_g X_g U^2_g] < \infty \label{eq:momentXU}. 
\end{equation}
Without loss of generality, we show the case with a scalar $X$. 
For \eqref{eq:momentXX}, the identical distribution of $Z_{gi}$ across $i$ (Assumption \ref{a:cluster}.\ref{a:cluster:rv}) and the independence between $N_g$ and $Z_{gi}$ (Assumption \ref{a:cluster}.\ref{a:cluster:indep}) imply
\begin{equation*}
\mathbb{E}[N_g^{-1} X'_g X_g] = \mathbb{E}[N_g^{-1}\sum^{N_g}_{i=1} X^2_{gi}] =  \mathbb{E}[X^2_{gi}] <  \infty. 
\end{equation*}
For \eqref{eq:momentXU}, we have
\begin{equation*}
\mathbb{E}[N_g^{-2} X'_g X_g U^2_g] = \mathbb{E}[N_g^{-2}\sum^{N_g}_{i=1} X^2_{gi}U^2_{gi}] \overset{(1)}=  \mathbb{E}[N_g^{-1}] \mathbb{E}[X^2_{gi}U^2_{gi}] \overset{(2)}<\mathbb{E}[X^2_{gi}U^2_{gi}] \overset{(3)}<  \infty, 
\end{equation*}
where (1) follows from the identical distribution of $Z_{gi}$ across $i$ (Assumption \ref{a:cluster}.\ref{a:cluster:rv}) and the independence between $N_g$ and $Z_{gi}$ (Assumption \ref{a:cluster}.\ref{a:cluster:indep}), 
(2) follows from the fact that $N_g \geq 1$ almost surely, and 
(3) follows from Assumption \ref{a:cluster}.\ref{a:cluster:rv} with $\alpha>2$. 

Now we show the consistency of the variance estimator, that is,
\begin{equation*}
G\widehat V_{\hat\theta}^{\text{WCR}} \xrightarrow{p} V^{\text{WCR}}.
\end{equation*}
The convergence $G^{-1}\sum_{g=1}^G N_g^{-1} X_g'X_g \overset{p}\rightarrow \mathbb{E}[N_g^{-1} X'_g X_g]$ follows from the law of large numbers given \eqref{eq:momentXX} and Assumption \ref{a:normal}.1. 
Given that $a_n \rightarrow 1$ almost surely, it remains to show that
\begin{equation*}
G^{-1}\sum_{g=1}^G N_g^{-2} \widehat S_g \widehat S_g' \overset{p}\rightarrow \mathbb{E}[N_g^{-2} X'_g X_g U^2_g]. 
\end{equation*}
To this end, we first write
\begin{align*}
\widehat S_g = &\, \sum_{i = 1}^{N_g} X_{gi} \widehat U_{gi} \\
 = &\, \sum_{i = 1}^{N_g} X_{gi}U_{gi} - \sum_{i = 1}^{N_g}X_{gi}^2 \, (\widehat\theta-\theta) \\
 = &\, S_g - M_g \, (\widehat\theta-\theta),
\end{align*}
where we again illustrate with the scalar $X$ case. 
Then, we have
\begin{align*}
& G^{-1}\sum_{g=1}^G N_g^{-2} \widehat S_g \widehat S_g' \\
= &\, G^{-1}\sum_{g=1}^G N_g^{-2}  S_g^2 - 2G^{-1}\sum_{g=1}^G N_g^{-2} S_g M_g (\widehat\theta-\theta) \\
 &\, + G^{-1}\sum_{g=1}^G N_g^{-2} M_g^2 (\widehat\theta-\theta)^2 \\
\overset{(1)} = &\, \mathbb{E}[N_g^{-2}  S_g^2] - 2\mathbb{E}[N_g^{-2} S_g M_g](\widehat\theta-\theta) +\mathbb{E}[N_g^{-2} M_g^2](\widehat\theta-\theta)^2 + o_p(1) \\
\overset{(2)} = &\, \mathbb{E}[N_g^{-2}  S_g^2] + o_p(1),
\end{align*}
where (1) follows from the law of large numbers given \eqref{eq:momentXX}, \eqref{eq:momentXU}, Assumption \ref{a:cluster}.\ref{a:cluster:rv} and Assumption \ref{a:normal}.1, and 
(2) follows from the consistency of $\widehat\theta$. 
\end{proof}

%%%%%%%%%%%%%%%%%%%%%%%%%%%%
\subsection{Proof of Proposition \ref{prop:jackknife}}\label{sec:prop:jackknife}
%%%%%%%%%%%%%%%%%%%%%%%%%%%%
\begin{proof}
Without loss of generality and for ease of writing, we consider the case with a scalar $X_{gi}$.
Let 
$
Z_{1g}=X_{g}^{\prime }X_{g}/N_{g}
$
and
$
Z_{2g}=X_{g}^{\prime }Y_{g}/N_{g}.
$
We can write
\begin{equation*}
\theta =g\left( \mathbb{E}\left[ Z_{1g}\right] ,\mathbb{E}\left[ Z_{2g}\right]\right) 
\end{equation*}
and
\begin{equation*}
\hat{\theta}^{\text{WCR}}=g\left( \bar{Z}_{1g},\bar{Z}_{2g}\right),
\end{equation*}%
where $g\left( a,b\right) =b/a$, $\bar{Z}_{1g}=G^{-1}\sum_{g=1}^{G}Z_{1g}$, and $\bar{Z}_{2g}=G^{-1}\sum_{g=1}^{G}Z_{2g}$. 
Let $\mu =\left( \mathbb{E}\left[ Z_{1g}\right] ,\mathbb{E}\left[ Z_{2g}\right] \right)$. 
Since $\bigtriangledown g$ is well-defined in a neighorbood of $\mu $, $\bigtriangledown g\left( \mu \right) \neq 0$, and $\bigtriangledown g$ is continuous at $\mu $, Theorem 2.1 of \citet{ShTu95} yields
\begin{equation*}
\frac{G\cdot \hat{V}_{\hat{\theta}}^{\text{WCR,JACK}}}{V^{\text{WCR}}}\rightarrow _{a.s.}1.
\end{equation*}
The statement of the proposition now follows.
\end{proof}

%%%%%%%%%%%%%%%%%%%%%%%%%%%%
\section{Testing the Robustness}\label{sec:test:appendix}
%%%%%%%%%%%%%%%%%%%%%%%%%%%%

As discussed in Section \ref{sec:pitfall}, essential for credible CR variance estimation under cluster sampling based on existing theory is a finite second moment of $S_g$.
Thus, we are interested in testing the hypotheses
\begin{align}\label{eq:hypothesis}
H_0: \mathbb{E}[||S_g||^2]<\infty
\end{align}

Note that a na\"ive application of common tail index or Pareto exponent estimators will not work to test this hypothesis, since the score $S_g$ in general involves unknown parameters that need to be estimated from data.
However, this hypothesis \eqref{eq:hypothesis} can be still formally tested by applying some existing theories and methods.
This appendix section briefly describes the testing procedure.

In general, we are interested in testing the hypothesis
\begin{align}\label{eq:hypothesis:appendix}
H_0: \mathbb{E}[||S_g||^r]<\infty
\qquad
\end{align}
It can be tested by applying a modified version of the method of \citet{SaWa22} suitably adapted to the cluster sampling framework.
A proposed testing procedure is outlined below.

Define $\widehat A^r_g = ({\widehat S_g}' {\widehat S_g})^{r/2}$ for $r \in \{1,2\}$.
Sort $\{\widehat A^r_g\}_{g=1}^G$ in the descending order into $\widehat A^r_{(1)} \ge \widehat A^r_{(2)} \ge \cdots \ge \widehat A^r_{(G)}$, i.e., $\widehat A^r_{(g)}$ denotes the $g$-th order statistic of $\{\widehat A^r_g\}_{g=1}^G$ for each $g \in \{1,\cdots,G\}$.
For a predetermined integer $k \ge 3$, collect the largest $k$ order statistics
$\widehat{\mathbf A}^r = (\widehat A^r_{(1)}, \cdots, A^r_{(k)})'$.
Implement the location- and scale-normalization of $\widehat{\mathbf A}^r$ to obtain
$$
\widehat{\mathbf A}^r_\ast = \frac{\widehat{\mathbf A}^r - \widehat A^r_{(k)}}{\widehat A^r_{(1)} - \widehat A^r_{(k)}}.
$$
We reject the null hypothesis \eqref{eq:hypothesis:appendix} of a finite $r$-th moment of $S_g$ if
$$
\frac{\int_1^2 f_{{\mathbf V}_\ast}(\widehat{\mathbf A}^r_\ast; \xi) dW(\xi)}{\int_0^1 f_{{\mathbf V}_\ast}(\widehat{\mathbf A}^r_\ast; \xi) d\Lambda(\xi)} > 1,
$$
where $W$ denotes the uniform probability measure, $\Lambda$ is a pre-determined measure that transforms the composite null into a simple one while subsuming the critical value for size control, and $f_{{\mathbf V}_\ast}$ is defined by
$$
f_{{\mathbf V}_\ast}({\mathbf v}_\ast;\xi) 
=
\Gamma(k) \int_0^\infty s^{k-2}
\exp\left( (-1-\xi^{-1}) \left( \log(1+\xi s) + \sum_{i=2}^{k-1} \log(1+\xi v_{\ast i} s) \right) \right) 
ds 
$$
with $\Gamma$ denoting the gamma function.
See \citet{SaWa22} for details.
The Stata command `\texttt{testout}' associated with this article implements this test.\footnote{This Stata command is available through SSC by the command line: \texttt{ssc install testout}.}

%%%%%%%%%%%%%%%%%%%%%%%%%%%%
\section{Details of Section \ref{sec:self_normalized}}\label{sec:appendix:self_normalized}
%%%%%%%%%%%%%%%%%%%%%%%%%%%%

Section \ref{sec:self_normalized} argues that the self-normalized central limit theorem (CLT) may or may not hold under our framework.
This appendix section presents details of this argument. 

For the estimand $\theta = \mathbb{E}[Y_{gi}]$ for simplicity, consider the estimator
\begin{equation*}
\hat{\theta}=\frac{\sum_{g=1}^{G}Z_{g}}{N},
\end{equation*}
where $Z_{g}=\sum_{i=1}^{N_{g}}Y_{gi}$ and $N=\sum_{g=1}^{G}N_{g}$. 
For simplicity, assume that $Y_{gi}$ is identically distributed with mean zero and variance one.
Also assume the cluster-sampling framework in which observations are independent across $g$. 
Let $\Omega _{N}$ denote the variance of $\sqrt{N}\hat{\theta}$, i.e., $\mathbb{E}[ N\hat{\theta}^{2}]$.

We now consider three cases of within-cluster dependence: (i) $Y_{gi}$ is i.i.d. across $i$ within each $g$ (i.e., no cluster dependence); (ii) $Y_{gi}=Y_{gj}$ for all $i$ and $j$ within the same cluster (i.e., the strongest form of cluster dependence); and (iii) a combination of the cases (i) and (ii).

\begin{description}
\item[Case (i)] 
Suppose that $Y_{gi}$ is i.i.d. across $i$.
The self-normalized CLT considers
\begin{equation*}
\left(\mathbb{E}[\hat{\theta}^{2}]\right)^{-1/2}\hat{\theta} \overset{d}{\rightarrow }\mathcal{N}\left( 0,1\right).
\end{equation*}
Since
$
\mathbb{E}[ N\hat{\theta}^{2} ] =\mathbb{E}\left[ Y_{gi}^{2}\right]=1
$
under the independence across $i$ and $g$, we have
\begin{equation}\label{eq:self_normalized}
\left(\mathbb{E}[\hat{\theta}^{2}]\right)^{-1/2}\hat{\theta}
=
\sqrt{N}\hat{\theta}=\frac{G^{-1/2}\sum_{g=1}^{G}Z_{g}}{\sqrt{\frac{1}{G}\sum_{g=1}^{G}N_{g}}}.
\end{equation}
By the law of large numbers and the assumption that $N_{g}$ is regularly
varying with exponent $\beta >1$, we have
\begin{equation*}
\frac{1}{G}\sum_{g=1}^{G}N_{g}\overset{d}{\rightarrow }\mathbb{E}\left[ N_{g}
\right] <\infty 
\end{equation*}
for the denominator of \eqref{eq:self_normalized}.
The independence within cluster implies that conditional on $\{N_g\}^G_{g=1}$, 
\begin{equation*}
N^{-1/2}\sum_{g=1}^{G}\sum_{i=1}^{N_{g}}Y_{gi}\overset{d}{\rightarrow }\mathcal{N}\left( 0,1\right) \text{.}
\end{equation*}
for the numerator of \eqref{eq:self_normalized}.
Note that the convergence rate is $N^{-1/2}$, instead of $G^{-1/2}$.
Therefore, the self-normalized CLT still holds, but with the convergence rate being $N^{-1/2}$, instead of $G^{-1/2}$ if we treat $\{N_g\}^G_{g=1}$ as fixed sequences of constants. 
Now consider $\{N_g\}^G_{g=1}$ as random variables. 
Given the Pareto tail of $N_{g}$, we have that
\begin{equation*}
N=\sum_{g=1}^{G}N_{g} =O_{p}\left( G\right).
\end{equation*}
It follows that $G^{-1/2}\sum_{g=1}^{G}\sum_{i=1}^{N_{g}}e_{gi}=O_{p}\left(1\right) $. 

\item[Case (ii)] Consider the case with perfect within-cluster dependence, i.e., $Y_{gi}\equiv Y_{g}$ for all $i\in \{1,...,N_{g}\}$ for each $g$.
In this case, $Z_{g}=\sum_{i=1}^{N_{g}}Y_{gi}=N_{g}Y_{g}$, yielding that
\begin{equation*}
\sqrt{N}\hat{\theta}=\frac{G^{-1/2}\sum_{g=1}^{G}N_{g}Y_{g}}{\sqrt{\frac{1}{G}\sum_{g=1}^{G}N_{g}}}.
\end{equation*}%
The denominator still converges to $\sqrt{\mathbb{E}\left[ N_{g}\right] }$.
For the numerator, since $N_{g}Y_{g}$ is i.i.d. across $g$ and the two factors are independent with regularly varying tails, \citet[][Proposition 1.3.9]{mikosch1999regular} implies that the product $N_{g}Y_{g}$ also has regularly varying tail with exponent $\beta <2$. 
Therefore, $G^{-1/2}\sum_{g=1}^{G}Z_{g} = G^{-1/2}\sum_{g=1}^{G}N_{g} Y_{g}$ is no longer $O_{p}\left( 1\right) $. 
More specifically, $\text{Var}[ G^{-1/2}\sum_{g=1}^{G}N_{g} Y_{g}]$ is equal to $\text{Var}[N_g]\cdot \text{Var}[Y_g]$, which is infinite given $\beta<2$. 
In fact, \citet[][Theorem 1]{geluk2000stable} implies that if the distribution of $N_gY_g$ is $\alpha$-stable, under some sequences of constants $a_G \simeq n^{1/\beta} \rightarrow \infty$ and $b_G \in \mathbb{R}$, the limit distribution function
\begin{equation*}
H(x) = \lim_{G \rightarrow \infty} \mathbb{P}\left( \frac{1}{a_{G}} \sum_{g=1}^G Z_g - b_G > x \right)
\end{equation*}
has the characteristic function
\begin{equation*}
\psi _{\beta}\left( s\right) 
=
\exp \left\{ -\left( \left\vert s\right\vert ^{\beta }+is \left( 1-\beta\right) \tan(\beta\pi/2) \frac{\left\vert s\right\vert ^{\beta -1}-1}{\beta -1}\right) \right\}. 
\end{equation*}
Thus, the CLT fails, and the asymptotic distribution will be non-Gaussian.
Therefore, even the jackknife standard error fails in this scenario. 
See, for example, Figures 5 and 6 in \citet{MaNiWe2022fast}.

\item[Case (iii)] Combining the above two cases, we now consider 
\begin{equation*}
Y_{gi}=\rho _{G}R_{g}+e_{gi},
\end{equation*}
where $R_{g}$ can be thought as a cluster-specific random effect and $e_{gi}$ is a random noise, which is i.i.d. across both $i$ and $g$. 
The normalizing constant $\rho _{G}$ determines the weights of $R_{g}$ in $Y_{gi}$. 
Under this setting, we have
\begin{align}
\sqrt{N}\hat{\theta} =&\frac{G^{-1/2}\sum_{g=1}^{G}Z_{g}}{\sqrt{\frac{1}{G}\sum_{g=1}^{G}N_{g}}} \notag
\\
=&\frac{G^{-1/2}\rho _{G}\sum_{g=1}^{G}N_{g}R_{g}}{\sqrt{\frac{1}{G} \sum_{g=1}^{G}N_{g}}}+\frac{G^{-1/2}\sum_{g=1}^{G}\sum_{i=1}^{N_{g}}e_{gi}}{\sqrt{\frac{1}{G}\sum_{g=1}^{G}N_{g}}}. \label{eq:two part}
\end{align}
Following the same arguments as those in Case (ii), the first item above is asymptotically non-Gaussian (after some suitable normalization), but the second term is asymptotically Gaussian. 
The orders of magnitudes of them depend on the distribution of $\left( R_{g},N_{g},e_{gi}\right) $. 
For example, if $\mathbb{E}\left[ R_{g} \right] =0 $ and $\mathbb{E}\left[ R_{g}^{2}\right] <\infty $, then $N_gR_g$ again has a regularly varying tail with exponent $\beta<2$ \citep[e.g.,][Theorem 3]{EmGo80}. 
The generalized central limit theorem \citep[e.g.,][Chapter 11]{Ib13} implies that
$
\sum_{g=1}^{G}N_{g}R_{g} \simeq_p G^{1/\beta }.
$
For the second term in \eqref{eq:two part}, Case (i) derives that $G^{-1/2}\sum_{g=1}^{G}\sum_{i=1}^{N_{g}}e_{gi}=O_{p}\left(1\right) $. 
The non-Gaussian part then dominates the Gaussian part if $\rho_G G^{1/\beta-1/2}\rightarrow\infty$ as $G\rightarrow\infty$. 
Since $\beta<2$, a constant $\rho_G$ will satisfy this condition. 
\end{description}

As a final remark, we note that Assumption 3 in \citet{DjMaNi19} could relax the condition on $N_g$ into $\sup_g N_g/N \rightarrow 0$ when the within-cluster dependence is strong. 
However, this assumption fails under our framework where $N_g$ is treated as a random variable. 
More specifically, consider Case (ii) again for illustration. 
Let $\mu_N$ denote the reciprocal of the variance of $\hat{\theta}$ conditional on $\{N_g\}^G_{g=1}$ as in \citet{DjMaNi19}. 
The above derivation yields that 
\begin{align*}
\text{Var}[\hat{\theta}|\{N_g\}^G_{g=1}] 
= &\, \frac{\sum^G_{g=1}N_g^2\text{Var}[Y_{gi}]}{(\sum^G_{g=1}N_g)^2} \\
= &\, \frac{\sum^G_{g=1}N_g^2\text{Var}[Y_{gi}]G^{-2}}{(G^{-1}\sum^G_{g=1}N_g)^2} \\
\simeq_p &\, G^{2/\beta-2},
\end{align*}
and hence $\mu_N \simeq_p G^{2-2/\beta}$. 
Therefore, for any constant $\lambda>0$, we have that 
\[
\mu_N^{\frac{2+\lambda}{2+2\lambda}}\frac{\sup_g N_g}{N} \simeq_p G^{\rho(\lambda)},
\]
where $\rho(\lambda) = (2-2/\beta)[(2+\lambda)/(2+2\lambda)-1/2]$. 
Recall that $\beta \in (1,2)$, yielding that $\rho(\lambda)>0$ for all $\lambda>0$.
Then the above term diverges with probability approaching one, suggesting that Assumption 3 in \citet{DjMaNi19} is violated. 
\end{document}